\documentclass[amsmath,amssymb,showpacs]{revtex4}
\usepackage{graphicx}
\usepackage{dcolumn}
\usepackage{bm}
\makeatletter \@addtoreset{equation}{section} \makeatother

\newcommand{\D}{\displaystyle}

\newcommand{\p}{\partial}
\newcommand{\e}{\epsilon}

\begin{document}


\title{Solitary Waves Bifurcated from Bloch Band Edges in Two-dimensional Periodic Media}
\author{Zuoqiang Shi$^{1}$, Jianke Yang$^{2}$}
\affiliation{%
$^{1}$Zhou Pei-Yuan Center for Applied Mathematics,
Tsinghua University, Beijing 100084, China \\
$^{2}$Department of Mathematics and Statistics, University of
Vermont, Burlington, VT 05401, USA
}%


\begin{abstract}
Solitary waves bifurcated from edges of Bloch bands in
two-dimensional periodic media are determined both analytically and
numerically in the context of a two-dimensional nonlinear
Schr\"odinger equation with a periodic potential. Using multi-scale
perturbation methods, envelope equations of solitary waves near
Bloch bands are analytically derived. These envelope equations
reveal that solitary waves can bifurcate from edges of Bloch bands
under either focusing or defocusing nonlinearity, depending on the
signs of second-order dispersion coefficients at the edge points.
Interestingly, at edge points with two linearly independent Bloch
modes, the envelope equations lead to a host of solitary wave
structures including reduced-symmetry solitons, dipole-array
solitons, vortex-cell solitons, and so on --- many of which have
never been reported before. It is also shown analytically that the
centers of envelope solutions can only be positioned at four
possible locations at or between potential peaks. Numerically,
families of these solitary waves are directly computed both near and
far away from band edges. Near the band edges, the numerical
solutions spread over many lattice sites, and they fully agree with
the analytical solutions obtained from envelope equations. Far away
from the band edges, solitary waves are strongly localized with
intensity and phase profiles characteristic of individual families.

\end{abstract}

\pacs{42.65.Tg, 05.45.Yv}
\maketitle


\section{Introduction}
Nonlinear wave propagation in periodic media is attracting a lot of
attention these days. This was stimulated in part by rapid advances
in optics, Bose-Einstein condensates, and related fields. In optics,
various periodic and quasi-periodic structures (such as photonic
crystals, photonic crystal fibers, periodic waveguide arrays and
photonic lattices) have been constructed by ingenious experimental
techniques, with applications to light routing, switching and
optical information processing
\cite{Joannopoulos,Russell,Eisenberg,ChenLattice,Segev03,Denz,ChenMartin,Stegeman}.
These periodic media create a wide range of new phenomena for light
propagation, even in the linear regime. For instance, the
diffraction of light in a periodic medium exhibits distinctively
different patterns from homogeneous diffraction \cite{Eisenberg}. If
the periodic medium has a local defect, this defect can guide light
by a totally new physical mechanism called repeated Bragg
reflections
\cite{Joannopoulos,Russell,Silberberg_defect,YangChenOL,YangChendefect,
MakazyukPRL,ChenExpress}. When the nonlinear effects become
significant, such as with high-power beams or strongly nonlinear
materials, the physical phenomena would be even richer and more
complex, and their understanding is far from complete yet. In
Bose-Einstein condensates, one direction of recent research is to
load the condensates into periodic optical lattices
\cite{Pitaevskii,BEC_lattice,Kivshar_Opt_Exp}. This problem and the
above nonlinear optics problems are closely related, and are often
analyzed together.

Solitary waves play an important role in nonlinear wave systems.
These waves are nonlinear localized structures which propagate
without change of shape. In physical communities, they are often
just called solitons, which we do occasionally in this paper as
well. In one-dimensional (1D) periodic media, solitary waves (called
lattice solitons) have been predicted and observed in optical
experiments
\cite{Christdoulides88,Eisenberg,Segev_1D,Neshev,Peli_1D,Stegeman}.
But in two and higher dimensions, periodic media can support a much
wider array of solitary wave structures, many of which have no
counterparts in 1D systems. So far, several types of 2D lattice
solitons in the semi-infinite bandgap (such as fundamental, dipole
and vortex solitons) as well as the first bandgap (such as
fundamental, vortex and reduced-symmetry gap solitons) have been
reported
\cite{Segev03,YangMuss,Malomed,Christdoulides03,Chen_vortex,Segev_vortex,
YangStudies,
Kivshar_Opt_Exp,Segev_highervortex,Kivshar_C_onemode,Kivshar_OE,Segev_random,Malomed_gap_vortex}.
Solitons in Bessel-ring lattices and 2D quasi-periodic lattices have
been reported as well \cite{Kartashov,Chen_ring,Ablowitz}. All these
works were either numerical or experimental, and an analytical
understanding on these solitons is still lacking. Some of these
solitons bifurcate from edges of Bloch bands. For instance, the gap
vortex solitons reported in \cite{Segev_highervortex} bifurcate from
the two X-symmetry points of the second Bloch band, and the
reduced-symmetry solitons reported in \cite{Kivshar_C_onemode}
bifurcate from a single X-symmetry point of the second Bloch band
(both under focusing nonlinearity). This raises the following
important questions: are there other types of solitary waves
bifurcated from Bloch bands in 2D periodic media? how can such
solitary waves be analytically predicted and classified?

In this paper, we determine all possible solitary-wave structures
bifurcated from edges of Bloch bands in 2D periodic media both
analytically and numerically, using the two-dimensional nonlinear
Schr\"odinger equation with a periodic potential as the mathematical
model. By multi-scale perturbation methods, we derive the envelope
equations of these solitary waves near band edges. We find that
these envelope equations admit solutions which lead to not only
solitons reported before (see
\cite{Segev_highervortex,Kivshar_C_onemode} for instance), but also
new solitary-wave structures such as dipole-array solitons
bifurcated from the second Bloch band under focusing nonlinearity,
and vortex-cell solitons bifurcated from the second Bloch band under
\emph{defocusing} nonlinearity. We also show analytically that the
centers of envelope solutions can only be positioned at four
possible locations at or between the potential peaks. We further
determine the whole families of these solitons both near and far
away from band edges directly using numerical methods
\cite{YangLakoba}. Near the band edges, the numerical solutions
spread over many lattice sites, and they fully agree with the
analytical solutions obtained from envelope equations. Far away from
the band edges, solitary waves are strongly localized, and their
intensity and phase profiles carry signatures of individual soliton
families. These studies provide a rather complete understanding of
solitary waves bifurcated from Bloch bands in 2D periodic media.

\section{The Mathematical Model}
The mathematical model we use for the study of solitary waves in 2D
periodic media is the 2D nonlinear Schr"odinger (NLS) equation with
a periodic potential:
\begin{eqnarray} \label{NLS}
iU_t+U_{xx}+U_{yy}-V(x,y)U+\sigma |U|^2U=0,
\end{eqnarray}
where $U(x, y, t)$ is a complex function, $\,V(x,y)\,$ is the
periodic potential (also called the lattice potential), and
$\sigma=\pm 1$ is the sign of nonlinearity. This model arises in
Bose-Einstein condensates trapped in a 2D optical lattice (where $t$
is time) \cite{Pitaevskii,Kivshar_Opt_Exp} as well as light
propagation in a periodic Kerr medium under paraxial approximation
(where $t$ is the distance of propagation). In certain optical
materials (such as photorefractive crystals), the nonlinearity is of
a different (saturable) type. But it is known that those different
types of nonlinearities give qualitatively similar results as the
Kerr nonlinearity above
\cite{YangMuss,Chen_vortex,Segev_vortex,YangNJP}.

In this article we take the lattice potential as
\begin{eqnarray} \label{V}
V(x,y)=V_0\left(\sin^2x+\sin^2y\right),
\end{eqnarray}
whose periods $L$ along the $x$ and $y$ directions are both equal to
$\pi$. This square-lattice potential can be readily engineered in
Bose-Einstein condensates \cite{BEC_lattice,Kivshar_Opt_Exp} and
optics \cite{Segev03,ChenMartin}. This potential is separable, which
makes our theoretical analysis a little easier. Similar analysis can
be repeated for other types of periodic potentials and
nonlinearities (such as saturable nonlinearities in photorefractive
crystals \cite{Segev03,YangNJP}) with minimal changes. Without loss
of generality, when we carry out specific computations, we always
set $V_0=6$ in the potential (\ref{V}).

Solitary waves in Eq. (\ref{NLS}) are sought in the form
\begin{equation}
U(x,y,t)=u(x,y)e^{-i\mu t},
\end{equation}
where amplitude function $u(x, y)$ is a solution of the following
equation:
\begin{eqnarray}  \label{u}
u_{xx}+u_{yy}-[F(x)+F(y)]u+\mu u+\sigma |u|^2u=0,
\end{eqnarray}
\begin{equation}
F(x)=V_0\sin^2x,
\end{equation}
and $\mu$ is a propagation constant.

In this article, we determine solitary waves in Eq. (\ref{u}) which
are bifurcated from the Bloch bands (i.e. continuous spectrum) of
that equation. To do so, information about the Bloch bands of Eq.
(\ref{u}) is essential. Such Bloch bands will be analyzed first
below.

\section{Bloch bands and band gaps} \label{blochbands}
\begin{figure}
\center
\includegraphics[width=0.45\textwidth]{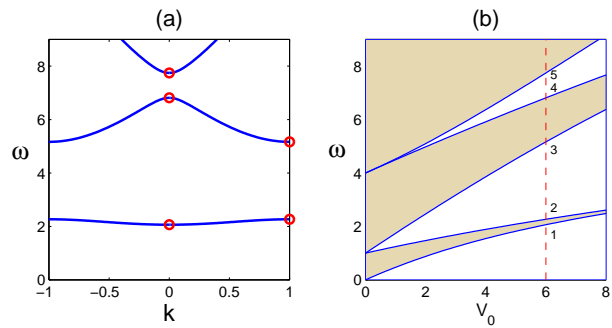}
\caption{(color online) (a) Dispersion curves of the 1D equation
(\ref{p}) with $V_0=6$; (b) Bloch bands (shaded regions) and
bandgaps at various values of potential levels $V_0$ in Eq.
(\ref{p}). The circle points in (a) correspond to edges of Bloch
bands marked by numbers 1-5 in (b). Bloch modes at these locations
are displayed in Fig. \ref{1Dbloch}. \label{1Ddispersion}}
\end{figure}

\begin{figure}
\center
\includegraphics[width=0.45\textwidth]{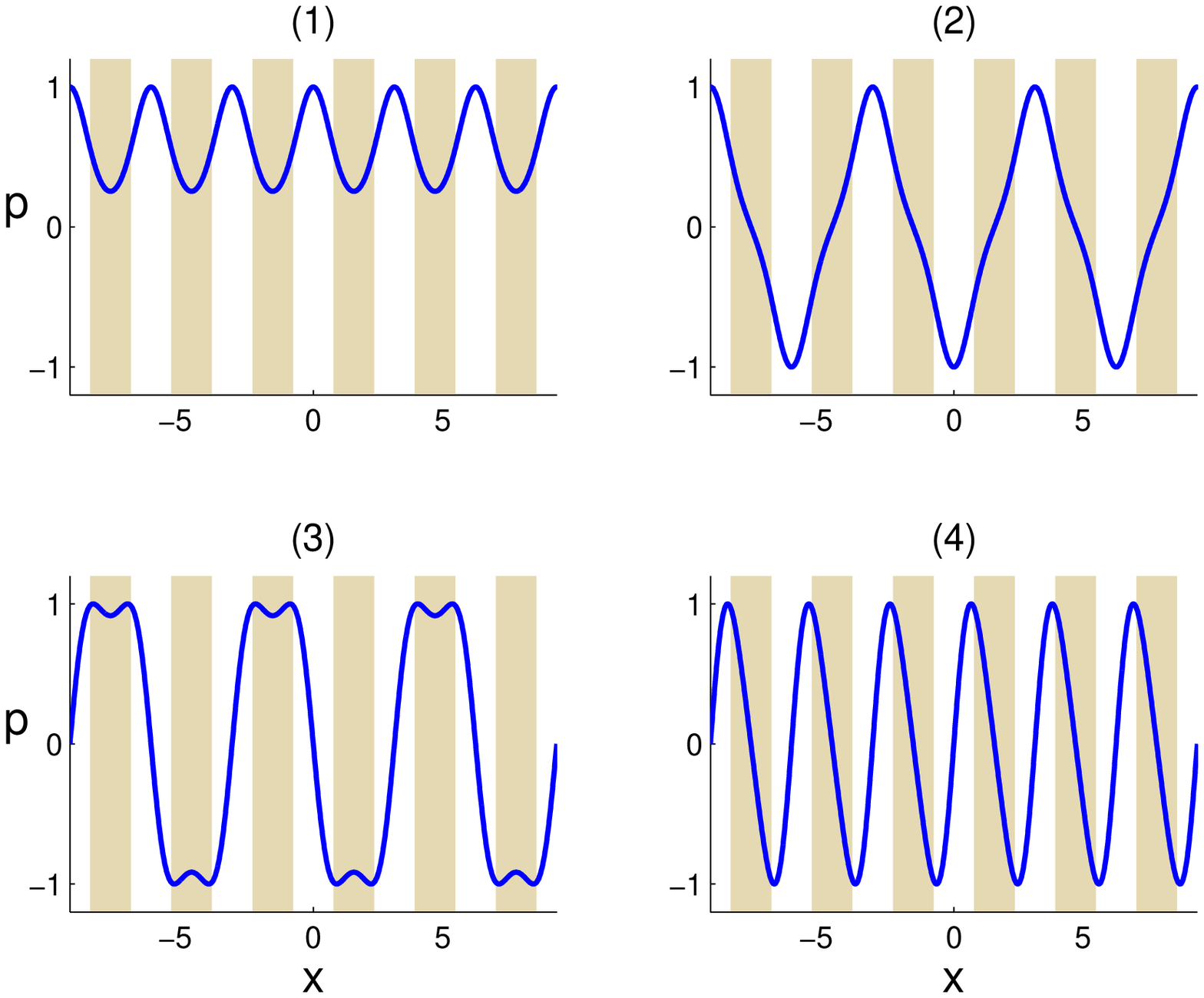}

\includegraphics[width=0.225\textwidth]{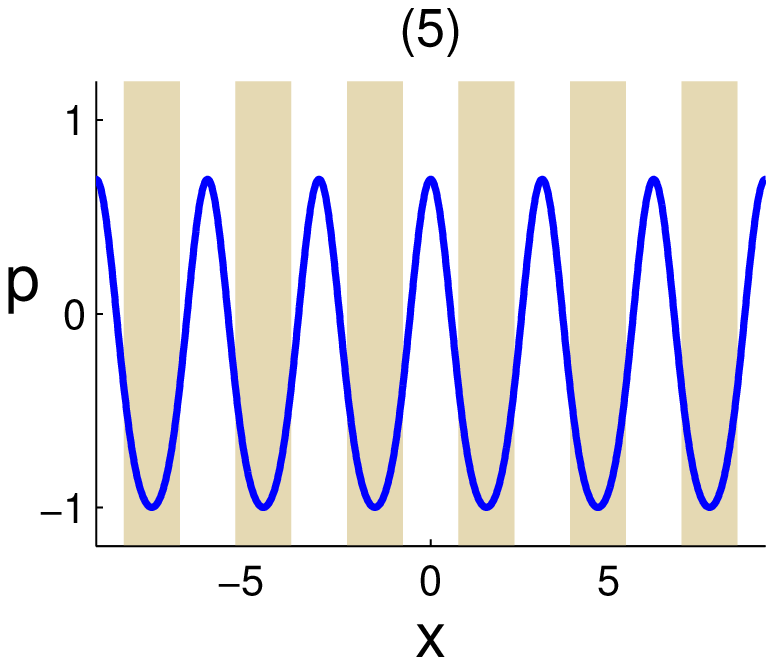}
\caption{One-dimensional Bloch waves of Eq. (\ref{p}) at the lowest
five edges of Bloch bands marked by circles in Fig.
\ref{1Ddispersion}(a) and by numbers in Fig. \ref{1Ddispersion}(b).
(1): $\omega_1=2.063182$; (2): $\omega_2=2.266735$; (3):
$\omega_3=5.165940$; (4): $\omega_4=6.81429$; (5)
$\omega_5=7.74678$. \label{1Dbloch}}
\end{figure}

When function $u(x, y)$ is infinitesimal, Eq. (\ref{u}) becomes a
linear equation:
\begin{eqnarray} \label{uF}
u_{xx}+u_{yy}-[F(x)+F(y)]u+\mu u=0.
\end{eqnarray}
Solutions of this linear equation are the Bloch modes, and the
corresponding propagation constants $\mu$ form Bloch bands. Since
the potential in (\ref{uF}) is separable, Bloch solutions and Bloch
bands of this 2D equation can be constructed from solutions of a 1D
equation. Specifically, the 2D Bloch solution $u(x, y)$ of Eq.
(\ref{uF}) and its propagation constant $\mu$ can be split into the
following form:
\begin{equation}  \label{connection}
u(x,y)=p(x; \omega_a)p(y; \omega_b), \quad \mu=\omega_a+\omega_b,
\end{equation}
where $p(x; \omega)$ is a solution of the following 1D equation:
\begin{equation}
p_{xx}-F(x)p+\omega p=0.   \label{p}
\end{equation}
This 1D equation is equivalent to the Mathieu's equation. Its
solution is
\begin{equation}
p(x; \omega)=e^{ikx}\tilde{p}(x; \omega),
\end{equation}
where $\tilde{p}(x; \omega)$ is periodic with the same period $\pi$
as the potential $F(x)$, and
$\omega=\omega(k)$
is the 1D dispersion relation. This dispersion diagram is shown in
Fig. \ref{1Ddispersion}(a) (for $V_0=6$). The bandgap structure of
this 1D equation (\ref{p}) at various values of $V_0$ is shown in
Fig. \ref{1Ddispersion}(b). Notice that in the 1D case, at any
non-zero value of $V_0$, bandgaps appear; in addition, the number of
bandgaps is infinite. The first five Bloch waves $p(x; \omega_k),
1\le k\le 5$ at the lowest five edges of Bloch bands
$\omega=\omega_k$ are displayed in Fig. \ref{1Dbloch}. These Bloch
waves have been normalized to have unit amplitude. Notice that these
solutions at band edges are all real-valued.

\begin{figure}
\center
\includegraphics[width=6cm, height=6cm]{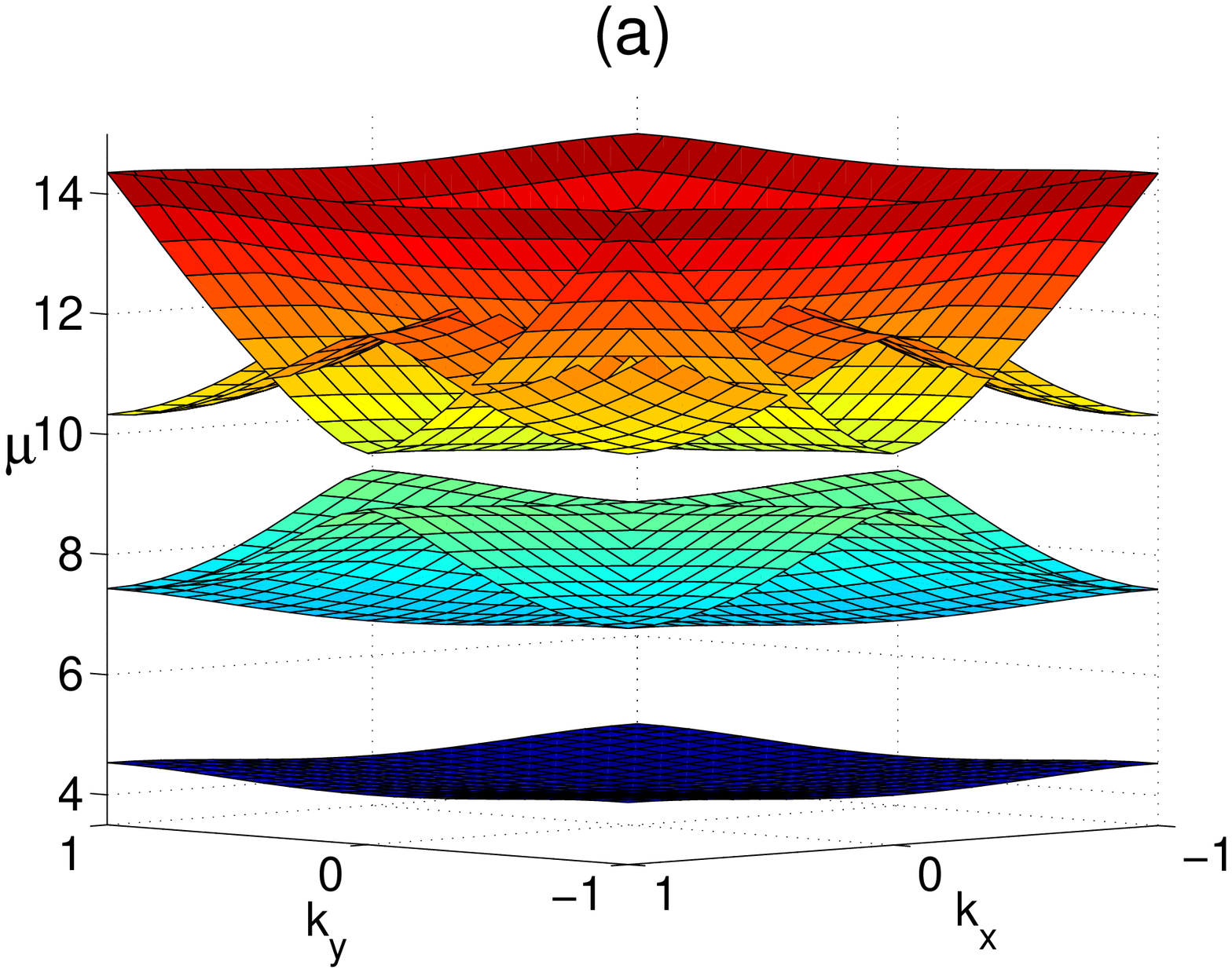}
\includegraphics[width=3cm, height=3cm]{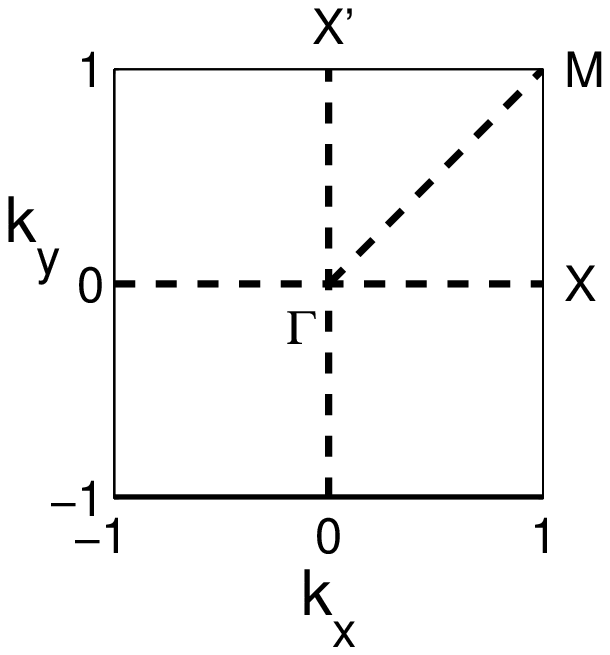}
\hspace{0.5cm}
\includegraphics[width=6cm, height=6cm]{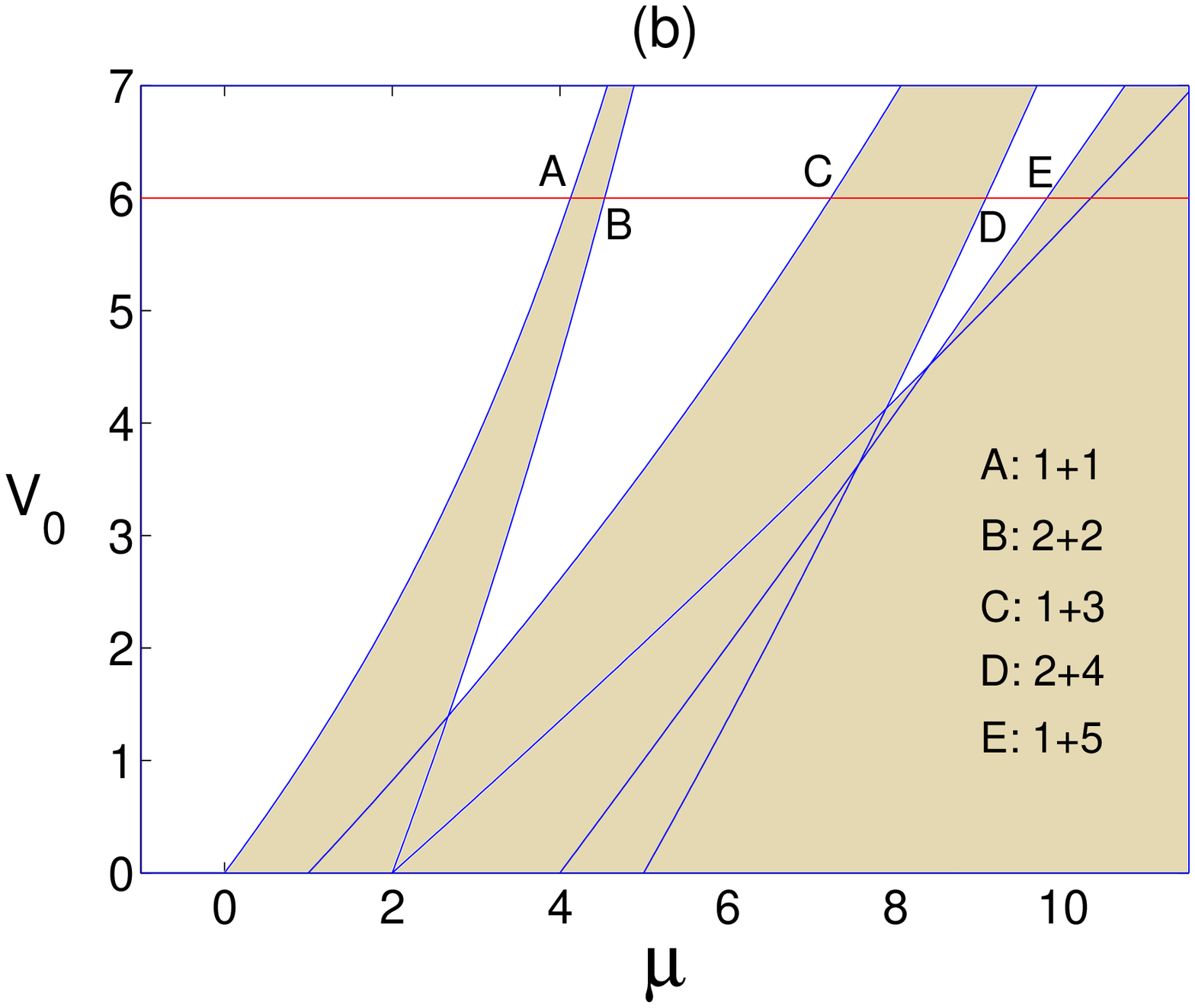}
\caption{(color online) (a) Dispersion surfaces of the 2D problem
(\ref{uF}) at $V_0=6$; the low figure at its side is the first
Brillouin zone; (b) the 2D bandgap structure for various values of
$V_0$. Letters 'A, B, C, D, E' mark the edges of Bloch bands at
$V_0=6$. Insets like 'A:1+1' are explained in the text.
\label{2Ddispersion}}
\end{figure}

Using these 1D dispersion results and the above connection between
1D and 2D Bloch solutions, we can construct the dispersion surfaces
and bandgap structures of the 2D problem (\ref{uF}). The 2D
Bloch-mode solution is of the form
\begin{equation}
u(x, y)=e^{ik_x \hspace{0.01cm} x+ik_y \hspace{0.01cm}
y}\tilde{p}[x; \omega(k_x)]\: \tilde{p}[y; \omega(k_y)],
\end{equation}
where
\begin{equation}
\mu=\omega(k_x)+\omega(k_y), \hspace{0.5cm} -1 \le k_x, k_y \le 1
\end{equation}
is the 2D dispersion relation, and $-1 \le k_x, k_y \le 1$ is the
first Brillouin zone. This 2D dispersion relation (at $V_0=6$) is
shown in Fig. \ref{2Ddispersion} (a). The 2D bandgap structure at
various values of $V_0$ is shown in Fig. \ref{2Ddispersion} (b).
Unlike the 1D case, for a given $V_0$ value, there are only a finite
number of bandgaps in the 2D problem. The first bandgap appears only
when $V_0>1.40$, the second bandgap appearing when $V_0 > 4.13$,
etc. As $V_0$ increases further, more bandgaps will be found. At the
potential strength $V_0=6$ as used in this paper, two bandgaps
exist. Edges of Bloch bands at this $V_0$ value are marked in Fig.
\ref{2Ddispersion} (b) as 'A, B, C, D, E' respectively.

Locations of Bloch-band edges in the first Brillouin zone are
important, as these locations reveal the symmetry properties of
Bloch modes. To clearly mark such locations, we plotted the first
Brillouin zone in Fig. \ref{2Ddispersion}. In the literature, the
center of this Brillouin zone is called the $\Gamma$ point, the four
corners called the $M$ points, and points $(k_x, k_y)=(1, 0), (0,
1)$ called the $X$ and $X'$ points respectively
\cite{Segev_highervortex}. These points are marked in the Brillouin
zone of Fig. \ref{2Ddispersion}. Note that the four corner ($M$)
points correspond to the same Bloch mode, but points $X$ and $X'$
lead to different (linearly independent) Bloch modes. In this first
Brillouin zone, Bloch-band edges A, B in Fig. \ref{2Ddispersion} (b)
are located at $\Gamma$ and $M$ symmetry points respectively. At
band edge C, there are two points on the dispersion surfaces which
are located at $X$ and $X'$ symmetry points. The same goes to the
band edge D. Band edge E is located at the $\Gamma$ symmetry point.

\begin{figure}
\center
\includegraphics[width=0.45\textwidth]{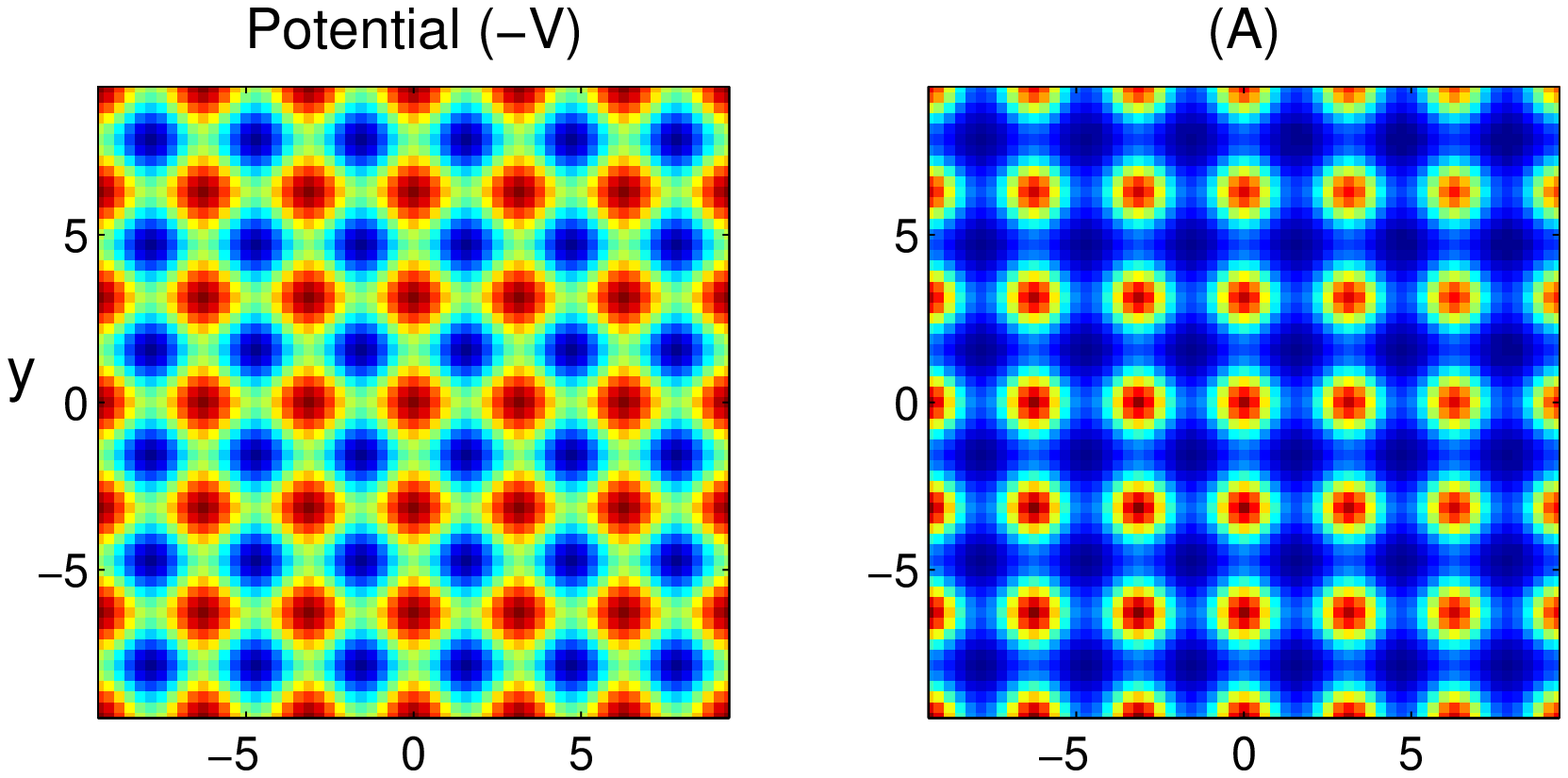}

\vspace{0.2cm}
\includegraphics[width=0.45\textwidth]{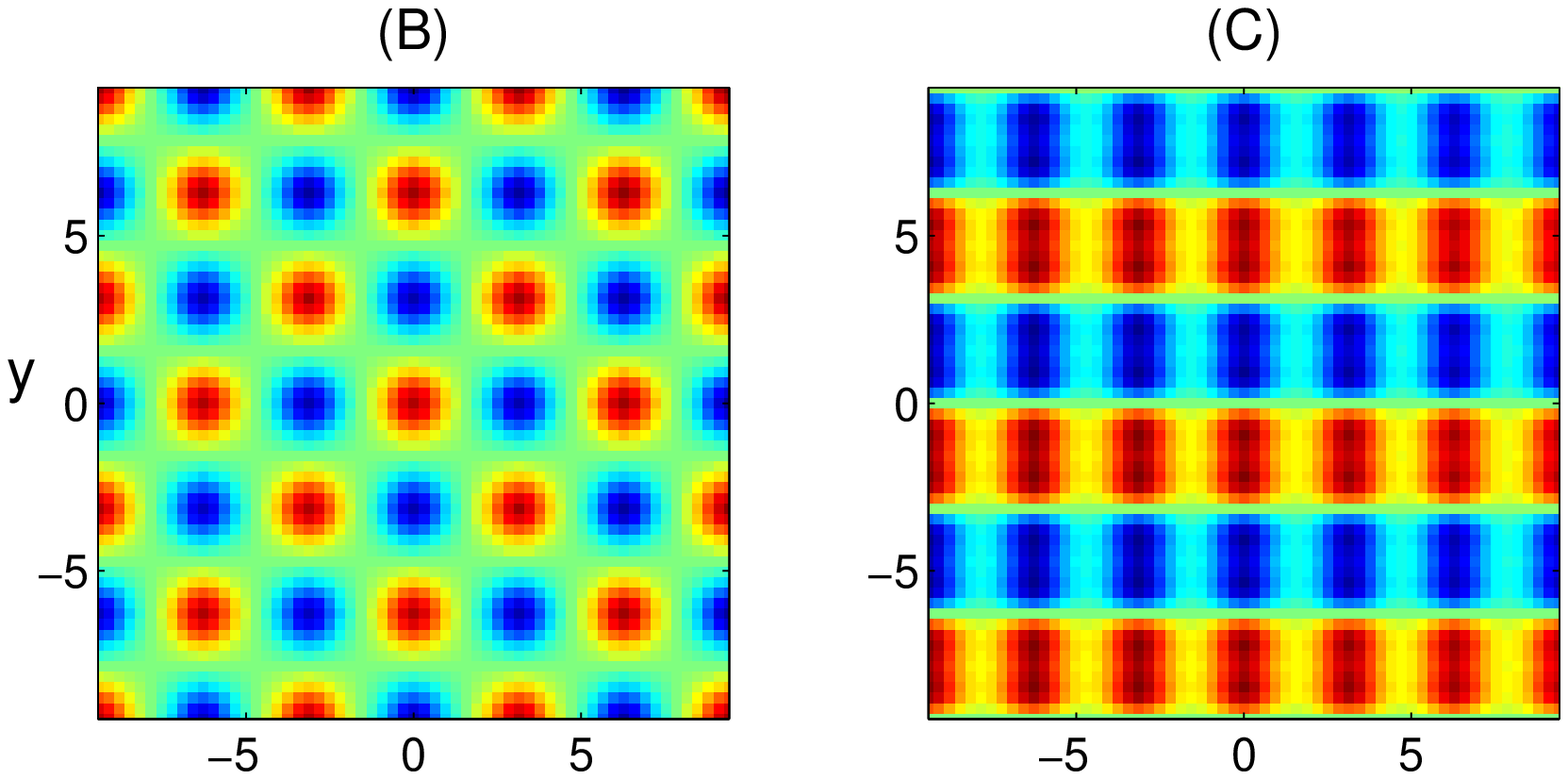}

\vspace{0.2cm}
\includegraphics[width=0.45\textwidth]{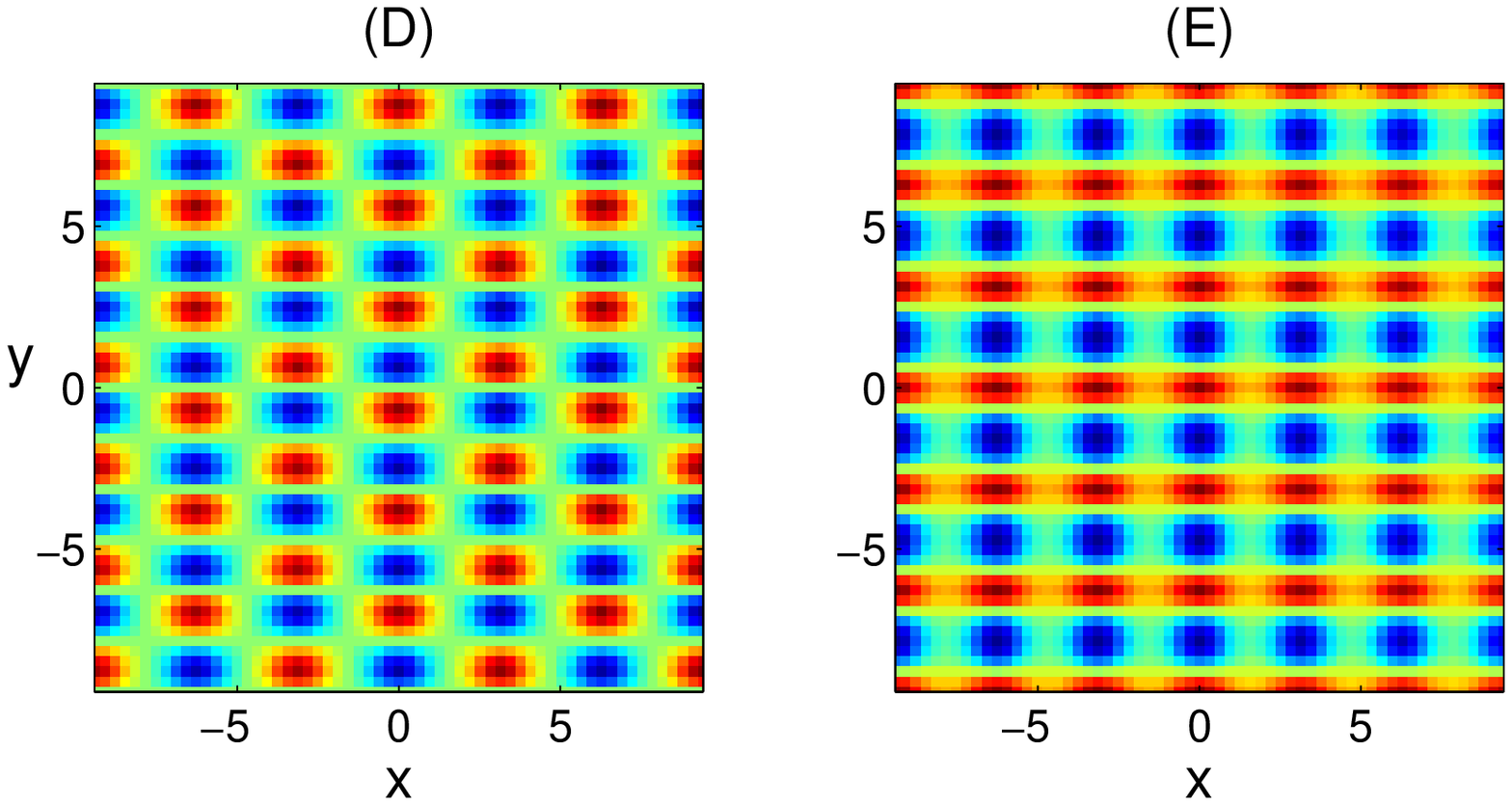}
\caption{(color online) Upper left: the potential function $-V(x,
y)$; (A, B, C, D, E) are Bloch modes at the edge points of Bloch
bands marked by the same letters in Fig. \ref{2Ddispersion}(b).
\label{Bloch_CD}}
\end{figure}

Now we examine 2D Bloch solutions at band edges. To illustrate, we
consider the edge points A, B, C, D and E in Fig. \ref{2Ddispersion}
(b), where $V_0=6$. At both points A and B, there is a {\it single}
Bloch solution. The Bloch solution at point A (located at $\Gamma$
point of the Brillouin zone) is $u(x, y)=p(x; \omega_1)p(y;
\omega_1)$, where $p(x; \omega_1)$ is shown in Fig.
\ref{1Dbloch}(1). This solution is displayed in Fig.
\ref{Bloch_CD}(A). Its propagation constant is $\mu=2\omega_1$.
Notice that this solution has the symmetry $u(x, y)=u(y, x)$. For
convenience, we denote point A as "$1+1$", and is so indicated in
Fig. \ref{2Ddispersion} (b). Similarly, the Bloch solution at point
B (located at $M$ point of the Brillouin zone) is $u(x, y)=p(x;
\omega_2)p(y; \omega_2)$, where $p(x; \omega_2)$ is shown in Fig.
\ref{1Dbloch}(2). This solution is displayed in Fig.
\ref{Bloch_CD}(B), and its propagation constant is $\mu=2\omega_2$.
This solution has the symmetry $u(x, y)=u(y, x)$ as well. Point B is
"$2+2$" in our notations. Points C, D, E are different from A and B
and are more interesting. At these points, there are {\it two}
linearly independent Bloch solutions, $u(x, y)$ and $u(y, x)$. At
point C, these two solutions are $u(x, y)=p(x; \omega_1)p(y;
\omega_3)$ and $u(y, x)=p(y; \omega_1)p(x; \omega_3)$, with the same
propagation constant $\mu=\omega_1+\omega_3$. Here $p(x; \omega_3)$
is shown in Fig. \ref{1Dbloch}(3). These solutions correspond to the
$X$ and $X'$ points in the Brillouin zone. Point C is thus "1+3".
Its $u(x, y)$ solution is displayed in Fig. \ref{Bloch_CD}(C); the
$u(y, x)$ solution is just a $90^\circ$ rotation of $u(x, y)$ in
Fig. \ref{Bloch_CD}(C) and thus not shown. Point D is "2+4", where
the two linearly independent Bloch solutions are $u(x, y)=p(x;
\omega_2)p(y; \omega_4)$ and $u(y, x)=p(y; \omega_2)p(x; \omega_4)$,
the former of which is displayed in Fig. \ref{Bloch_CD}(D). These
solutions also correspond to the $X$ and $X'$ points in the
Brillouin zone. Point E is "1+5", and the two Bloch solutions are
$u(x, y)=p(x; \omega_1)p(y; \omega_5)$ and $u(y, x)=p(y;
\omega_1)p(x; \omega_5)$ (the former is shown in Fig.
\ref{Bloch_CD}(E)). These solutions correspond to the $\Gamma$
symmetry point in the Brillouin zone. It is interesting that at
point E, {\it the two different Bloch modes are both located at the
same $\Gamma$ point}, while at points C and D, their two Bloch modes
are located at different symmetry points ($X$ and $X'$) in the
Brillouin zone. This difference between points E and \{C, D\} will
manifest itself later in solitary wave bifurcations. Because of the
existence of two linearly independent Bloch solutions at band edges
C, D and E, their linear superpositions remain a solution. These
superpositions can give rise to interesting solution patterns, some
of which (such as vortex arrays) have been pointed out before
\cite{Segev_highervortex,MakazyukPRL}, while many others are new.
Solitary waves bifurcated from these linearly superimposed Bloch
modes are one of the subjects we will focus on in the remainder of
the paper.

The above Bloch solutions exist on band edges with infinitesimal
amplitudes. When amplitudes of these solutions increase, these Bloch
solutions may localize and form solitary-wave structures. The
corresponding propagation constant $\mu$ then moves from band edges
into band gaps. In the next section, we analyze how solitary waves
bifurcate from Bloch solutions of band edges using multi-scale
perturbation methods.

\section{Envelope equations of Bloch modes}
\label{envelope_derivation}
In this section, we develop an
asymptotic theory to analyze small-amplitude solitary waves
bifurcating from Bloch waves near band edges in Eq. (\ref{u}), and
derive their envelope equations. We have known that at an edge of a
Bloch band, there may be either one or two linearly independent
Bloch modes. The bifurcation analysis for these two cases are
similar, thus we will present detailed calculations only for the
latter case (which is a little more complex), and just give the
results for the former case.

\subsection{Derivation of envelope equations}
Let us consider a 2D band edge $\mu_0=\omega_{0,1}+\omega_{0,2}$,
where $\omega_{0, n}\; (n=1, 2)$ are 1D band edges, $\omega_{0,1}\ne
\omega_{0,2}$, and the two linearly independent Bloch modes are
$p_1(x)p_2(y)$ and $p_1(y)p_2(x)$ with $p_n(x)=p(x; \omega_{0, n})$.
Notice that
\begin{equation} \label{pkL}
p_n(x+L)=\pm p_n(x)
\end{equation}
since $\omega_{0,n}$ is a 1D band edge. Here $L=\pi$ is the period
of the 1D potential $F(x)$. When the solution $u(x, y)$ of Eq.
(\ref{u}) is infinitesimally small, this solution on the band edge
is then a linear superposition of these two Bloch modes in the
general case. When $\,u(x,y)\,$ is small but not infinitesimal, we
can expand the solution $u(x, y)$ of (\ref{u}) into a multi-scale
perturbation series:
\begin{eqnarray} \label{uexpand}
u=\e u_0+\e^2u_1+\e^3u_2+\cdots,
\end{eqnarray}
\begin{equation}  \label{muexpand}
\mu=\mu_0+\eta \e^2,
\end{equation}
where
\begin{eqnarray} \label{u0}
u_0=A_1(X,Y)p_1(x)p_2(y)+A_2(X,Y)p_2(x)p_1(y),
\end{eqnarray}
$\eta=\pm 1$, and $\, X=\e x, Y=\e y$ are long spatial scales of
envelope functions $A_1$ and $A_2$. Substituting the above
expansions into Eq. (\ref{u}), the equation at $O(\e)$ is
automatically satisfied. At order $O(\e^2)$, the equation for $u_1$
is
\begin{equation} \label{u1}
u_{1xx}+u_{1yy}-\left[F(x)+F(y)\right]
u_1+\mu_0u_1=-2\left(\frac{\p^2u_0}{\p x\p X}+\frac{\p^2u_0}{\p y\p
Y}\right).
\end{equation}
Its homogeneous equation has two linearly independent solutions,
$p_1(x)p_2(y)$, and $p_1(y)p_2(x)$. In order for the inhomogeneous
equation (\ref{u1}) to admit a solution, the following Fredholm
conditions
\begin{eqnarray}
\int_0^{2L}\int_0^{2L}\left(\frac{\p^2u_0}{\p x\p
X}+\frac{\p^2u_0}{\p y\p
Y}\right)p_1(x)p_2(y)\;dxdy=0,\\
\int_0^{2L}\int_0^{2L}\left(\frac{\p^2u_0}{\p x\p
X}+\frac{\p^2u_0}{\p y\p Y}\right)p_2(x)p_1(y)\;dxdy=0
\end{eqnarray}
must be satisfied. Here the integration length is $2L$ rather than
$L$ since the homogeneous solutions $p_1(x)p_2(y)$ and
$p_2(x)p_1(y)$ may have periods $2L$ along the $x$ and $y$
directions (see Eq. (\ref{pkL})). Recalling the $u_0$ solution
(\ref{u0}), it is easy to check that the above Fredholm conditions
are indeed satisfied automatically, thus we can find a solution for
Eq. (\ref{u1}) as
\begin{eqnarray} \label{u1_solution}
u_1=\frac{\p A_1}{\p X}\nu_1(x)p_2(y)+\frac{\p A_1}{\p
X}\nu_2(y)p_1(x)+\frac{\p A_2}{\p X}\nu_2(x)p_1(y)+\frac{\p A_2}{\p
X}\nu_1(y)p_2(x),
\end{eqnarray}
where $\nu_n(x)$ is a periodic solution of equation
\begin{eqnarray}  \label{nu}
\nu_{n, xx}-F(x)\nu_n+\omega_{0,n}\nu_n=-2p_{n,x}, \quad n=1, 2.
\end{eqnarray}
At $O(\epsilon^3)$, the equation for $u_2$ is
\begin{widetext}
\begin{eqnarray}
u_{2xx}+u_{2yy}-\left[F(x)+F(y)\right]u_2+\mu_0u_2=-\left(2\frac{\p^2u_1}{\p
x\p X}+2\frac{\p^2u_1}{\p y\p Y}+\frac{\p^2u_0}{\p
X^2}+\frac{\p^2u_0}{\p Y^2}+\eta u_0+|u_0|^2u_0\right).
\end{eqnarray}
\end{widetext}
Substituting the formulas (\ref{u0}) and (\ref{u1_solution}) for
$u_0$ and $u_1$ into this equation, we get
\begin{widetext}
\begin{eqnarray} \label{u2}
&&-\left\{u_{2xx}+u_{2yy}-\left[F(x)+F(y)\right] u_2+\mu_0u_2\right\}=\nonumber\\
\nonumber\\
&&\frac{\p^2 A_1}{\p X^2}\left[2\nu_1'(x)+p_1(x)\right]p_2(y)
+\frac{\p^2 A_1}{\p Y^2}\left[2\nu_2'(y)+p_2(y)\right]p_1(x)
\nonumber\\
&&+\frac{\p^2 A_2}{\p X^2}\left[2\nu_2'(x)+p_2(x)\right]p_1(y)
+\frac{\p^2 A_2}{\p Y^2}\left[2\nu_1'(y)+p_1(y)\right]p_2(x)
\nonumber\\
&&+2\frac{\p^2A_1}{\p X\p
Y}\left[p_1'(x)\nu_2(y)+p_2'(y)\nu_1(x)\right]+2\frac{\p^2A_2}{\p
X\p Y}\left[p_2'(x)\nu_1(y)+p_1'(y)\nu_2(x)\right]\nonumber\\
&&+p_1^3(x)p_2^3(y)|A_1|^2A_1+p_1(x)p_2(y)p_1^2(y)p_2^2(x)(\bar{A}_1A_2^2+2A_1|A_2|^2)\nonumber\\
&&+p_2^3(x)p_1^3(y)|A_2|^2A_2+p_1^2(x)p_2^2(y)p_1(y)p_2(x)(\bar{A}_2A_1^2+2A_2|A_1|^2)\nonumber\\
&&+\eta A_1(X,Y)p_1(x)p_2(y)+\eta A_2(X,Y)p_2(x)p_1(y).
\end{eqnarray}
\end{widetext}
Here the overbar represents complex conjugation. Before applying the
Fredholm conditions to this inhomogeneous equation, we notice the
following identities:
\begin{equation} \label{iden1}
\int_0^{2L} p_1(x)p_2(x) dx =0,
\end{equation}
\begin{equation} \label{iden2}
\int_0^{2L} p_1(x)\nu_2(x) dx=\int_0^{2L} p_2(x)\nu_1(x) dx,
\end{equation}
and
\begin{equation} \label{iden3} \int_0^{2L} \left[2\nu_n'(x)+p_n(x)\right]p_n(x) dx = D_n \int_0^{2L} p_n^2(x) dx, \quad n=1, 2,
\end{equation}
where
\begin{equation}  \label{Dn}
D_n \equiv \left. \frac{1}{2}\frac{d^2\omega}{dk^2}
\right|_{\omega=\omega_{0, n}}.
\end{equation}
Identity (\ref{iden1}) holds since $p_1(x)$ and $p_2(x)$ are the
eigenfunctions of the self-adjoint linear Schr\"odinger operator
with different eigenvalues. Identity (\ref{iden2}) can be confirmed
by taking the inner product between Eq. (\ref{nu}) and functions
$p_n(x)$. Identity (\ref{iden3}) can be verified by expanding the
solution of Eq. (\ref{p}) around the edge of the Bloch band
$\omega=\omega_{0, n}$ (see Eq. (15) in \cite{Peli_1D}). Utilizing
these identities and (\ref{pkL}), the Fredholm conditions for Eq.
(\ref{u2}) finally lead to the following coupled nonlinear equations
for the envelope functions $A_1$ and $A_2$:
\begin{widetext}
\begin{eqnarray}
D_1 \frac{\p^2A_1}{\p X^2}+D_2 \frac{\p^2A_1}{\p Y^2}+\eta A_1+
\sigma \left[\alpha |A_1|^2A_1+\beta \left(\bar{A}_1A_2^2+2
A_1|A_2|^2\right)+\gamma\left(|A_2|^2A_2+\bar{A}_2A_1^2+2A_2|A_1|^2\right)\right]=0,  \label{A}\\
D_2 \frac{\p^2A_2}{\p X^2}+D_1\frac{\p^2A_2}{\p Y^2}+\eta A_2 +
\sigma \left[\alpha|A_2|^2A_2+\beta\left(\bar{A}_2A_1^2+2
A_2|A_1|^2\right)+\gamma\left(|A_1|^2A_1+\bar{A}_1A_2^2+2A_1|A_2|^2\right)\right]=0.
\label{B}
\end{eqnarray}
\end{widetext}
Here
\begin{equation} \label{alpha}
\alpha=\frac{\D\int_0^{2L}\int_0^{2L}p_1^{\,4}(x)p_{\,2}^{\,4}(y)\;dxdy}{\D\int_0^{2L}\int_0^{2L}p_1^{\,2}(x)p_{\,2}^{\,2}(y)\;dxdy},
\end{equation}
\begin{equation} \label{beta}
\beta=\frac{\D\int_0^{2L}\int_0^{2L}p_1^{\,2}(x)p_{\,2}^{\,2}(x)p_1^{\,2}(y)p_{\,2}^{\,2}(y)\;dxdy}{\D\int_0^{2L}\int_0^{2L}p_1^{\,2}(x)p_{\,2}^{\,2}(y)\;dxdy},
\end{equation}
and
\begin{equation} \label{gamma}
\gamma=\frac{\D\int_0^{2L}\int_0^{2L}p_1^{\,3}(x)p_{\,2}(x)p_{\,2}^{\,3}(y)p_1(y)\;dxdy}{\D\int_0^{2L}\int_0^{2L}p_1^{\,2}(x)p_{\,2}^{\,2}(y)\;dxdy}.
\end{equation}
Notice that $\alpha$ and $\beta$ are always positive, but $\gamma$
may be positive, negative, or zero.

The coefficients in Eqs. (\ref{A})-(\ref{B}) can be readily
determined from Bloch solutions of the 1D equation (\ref{p}). In
particular, when the 1D Bloch waves $p_1(x)$ and $p_2(x)$ have been
normalized to have unit amplitude (see Fig. 2), we find that at
point C,
\begin{eqnarray} \label{Cdata}
D_1=0.434845,\quad D_2=2.422196, \quad \alpha=0.5821, \quad
\beta=0.1325, \quad \gamma=0;
\end{eqnarray}
at point D,
\begin{eqnarray} \label{Ddata}
D_1=-0.586799, \quad D_2= -13.264815, \quad \alpha=0.5256, \quad
\beta=0.1811, \quad \gamma=0;
\end{eqnarray}
and at point E,
\begin{eqnarray} \label{Edata}
D_1= 0.434845, \; D_2= 15.793172, \; \alpha=0.4684, \; \beta=0.0781,
\; \gamma=-0.0261.
\end{eqnarray}

At band edges where a single Bloch mode exists (such as points $A$
and $B$ in Fig. \ref{2Ddispersion}(b)), the envelope equation for
this single Bloch mode can be more easily derived. In this case,
this single Bloch mode must be of the form $p_1(x)p_1(y)$, where
$p_1(x)=p(x; \omega_{0,1})$, and $\omega_{0,1}$ is a band edge in
the 1D problem (\ref{p}). Unlike the previous case, interchanging
$x$ and $y$ of this Bloch solution does not give different Bloch
modes. The leading order solution $u_0(x, y)$ in the expansion
(\ref{uexpand}) now is $A_1(X, Y)p_1(x)p_1(y)$, and the envelope
equation for $A_1(X, Y)$ can be found to be
\begin{equation} \label{A_single}
D_1 \left(\frac{\p^2A_1}{\p X^2}+ \frac{\p^2A_1}{\p Y^2}\right)+\eta
A_1+ \sigma \alpha_0 |A_1|^2A_1=0,
\end{equation}
where $D_1$ is as given in Eq. (\ref{Dn}), and
\begin{equation} \label{alpha0}
\alpha_0=\frac{\D\int_0^{2L}\int_0^{2L}p_1^{\,4}(x)p_{\,1}^{\,4}(y)\;dxdy}
{\D\int_0^{2L}\int_0^{2L}p_1^{\,2}(x)p_{\,1}^{\,2}(y)\;dxdy}.
\end{equation}
In particular, when the Bloch solution $p_1(x)$ is normalized to
have unit amplitude, then at point A,
\begin{eqnarray} \label{Acoef}
D_1=0.434845,  \quad \alpha_0 = 0.462815;
\end{eqnarray}
and at point B,
\begin{eqnarray} \label{Bcoef}
D_1=-0.588073,  \quad \alpha_0 = 0.500922.
\end{eqnarray}
We note that for this single-Bloch-mode case, an envelope equation
similar to (\ref{A_single}) has been derived before in
\cite{Baizakov_envelope}.

\subsection{Locations of envelope solitons}
The envelope equations (\ref{A})-(\ref{B}) and (\ref{A_single}) are
translation-invariant. For instance, if $[A_1(X, Y), A_2(X, Y)]$ is
a solution of Eqs. (\ref{A})-(\ref{B}), so is $[A_1(X-X_0, Y-Y_0),
A_2(X-X_0, Y-Y_0)]$ where $(X_0, Y_0)$ are any constants. However,
only when $(X_0, Y_0)$ take some special values can the perturbation
series solution (\ref{uexpand}) truly satisfy the original equation
(\ref{u}). The reason is that $(X_0, Y_0)$ must satisfy certain
additional constraints. These constraints are exponentially small in
$\epsilon$, thus they can not be captured in the power series
expansions of (\ref{uexpand}), but need to be calculated using
asymptotics beyond all orders techniques \cite{Akylas} or other
equivalent methods \cite{Peli_1D}. In 1D problems, it has been shown
that envelope solitons can only be located at two positions relative
to the periodic lattice \cite{Peli_1D} (or the underlying periodic
wave train \cite{Akylas}). In the present 2D problem, we will show
below that envelope solitons can only be located at four positions
relative to the 2D periodic lattice by a method similar to that used
in \cite{Peli_1D}. As we have done in the previous subsection,
detailed derivations will be presented for the $(A_1, A_2)$
solutions of Eqs. (\ref{A})-(\ref{B}). Similar results hold for the
$A_1$ solution of Eq. (\ref{A_single}) as well.

First we derive two constraints for the envelope solutions.
Multiplying Eq (\ref{u}) by $\bar{u}_x$ or $\bar{u}_y$, adding its
conjugate equation, and integrating from $-\infty$ to $+\infty$, we
get the following two constraints:
\begin{eqnarray}
\int_{-\infty}^{+\infty}\int_{-\infty}^{+\infty}F'(x)|u(x,y)|^2dxdy=0, \\
\int_{-\infty}^{+\infty}\int_{-\infty}^{+\infty}F'(y)|u(x,y)|^2dxdy=0.
\end{eqnarray}
Substituting the perturbation expansion (\ref{uexpand}) of the
solution $u(x, y)$ into the above equations, these constraints at
the leading order become
\begin{eqnarray}
I_1(x_0,y_0)=\epsilon^2\int_{-\infty}^{+\infty}\int_{-\infty}^{+\infty}F'(x)|A_1
p_1(x)p_2(y)+A_2 p_2(x)p_1(y)|^2dxdy=0,  \label{constraint1}\\
I_2(x_0,y_0)=\epsilon^2\int_{-\infty}^{+\infty}\int_{-\infty}^{+\infty}F'(y)|A_1
p_1(x)p_2(y)+A_2 p_2(x)p_1(y)|^2dxdy=0. \label{constraint2}
\end{eqnarray}
Here
\begin{equation} \label{ABX0Y0}
A_k=A_k(X-X_0, Y-Y_0), \quad k=1, 2,
\end{equation}
and $(X_0, Y_0)=(\epsilon x_0, \epsilon y_0)$ is the center position
of the envelope solution $(A_1, A_2)$. In this paper, we consider
envelope functions $(A_1, A_2)$ such that $|A_1|^2, |A_2|^2$ and
$A_1\bar{A}_2+\bar{A}_1A_2$ are symmetric in $X$ and $Y$ about the
center position $(X_0, Y_0)$, i.e.
\begin{equation} \label{assum_A}
|A_k(-X,Y)|^2=|A_k(X,Y)|^2=|A_k(X,-Y)|^2, \quad k=1, 2,
\end{equation}
and similar relations hold for $A_1\bar{A}_2+\bar{A}_1A_2$. All
solitary-wave solutions of Eqs. (\ref{A})-(\ref{B}) that we know
satisfy these assumptions (up to spatial translations).

Now we examine the constraint (\ref{constraint1}). This constraint
can be rewritten as
\begin{eqnarray} \label{I1}
I_1(x_0,y_0)=\epsilon^2\int_{-\infty}^{+\infty}\int_{-\infty}^{+\infty}F'(x)
\left[|A_1|^2p_1^2(x)p_2^2(y)+|A_2|^2p_2^2(x)p_1^2(y)\right.\nonumber\\
+\left.(A_1\bar{A}_2+\bar{A}_1A_2)p_1(x)p_2(x)p_1(y)p_2(y)\right]dxdy=0.
\end{eqnarray}
Since $F'(x)$ is anti-symmetric, and $p_1^2(x), p_2^2(x)$ both
symmetric, functions $F'(x)p_1^2(x)p_2^2(y)$,
$F'(x)p_2^2(x)p_1^2(y)$ and $F'(x)p_1(x)p_2(x)p_1(y)p_2(y)$ have the
following Fourier series expansions:
\begin{equation}
F'(x)p_1^2(x)p_2^2(y)=\sum_{m=1}^{\infty}\sum_{n=0}^{\infty}
c^{(1)}_{m,n}\sin \frac{2\pi mx}{L} \cos \frac{2\pi ny}{L},
\end{equation}
\begin{equation}
F'(x)p_2^2(x)p_1^2(y)=\sum_{m=1}^{\infty}\sum_{n=0}^{\infty}
c^{(2)}_{m,n}\sin \frac{2\pi mx}{L} \cos \frac{2\pi ny}{L},
\end{equation}
\begin{equation}
F'(x)p_1(x)p_2(x)p_1(y)p_2(y)=\sum_{m=1}^{\infty}\sum_{n=0}^{\infty}
c^{(3)}_{m,n}\sin \frac{2\pi mx}{L} \cos \frac{2\pi ny}{L} +
\sum_{m=0}^{\infty}\sum_{n=1}^{\infty} d^{(3)}_{m,n}\cos \frac{2\pi
mx}{L} \sin \frac{2\pi ny}{L}.
\end{equation}
Here $d^{(3)}_{m,n}=0$ or $c^{(3)}_{m,n}=0$ (for all $m, n$) if
$p_1(x)p_2(x)$ is even or odd respectively. When the above Fourier
expansions are substituted into Eq. (\ref{I1}), every Fourier mode
in these expansions leads to an exponentially small term in
(\ref{I1}), and the exponential rate of decay of these terms is
larger for higher values of $m+n$. Keeping only the leading-order
term obtained from Fourier modes with $m+n=1$, Eq. (\ref{I1})
becomes
\begin{eqnarray}
I_1(x_0,y_0) =
\epsilon^2\int_{-\infty}^{+\infty}\int_{-\infty}^{+\infty}\left\{
\left[c_{1,0}^{(1)} |A_1|^2+c_{1,0}^{(2)} |A_2|^2 +
c_{1,0}^{(3)}(A_1\bar{A}_2+\bar{A}_1A_2)\right] \sin \frac{2\pi
x}{L}
\right. \nonumber \\
\left. +d_{1,0}^{(3)}(A_1\bar{A}_2+\bar{A}_1A_2)\sin \frac{2\pi
y}{L} \right\}dx dy.
\end{eqnarray}
Recalling Eq. (\ref{ABX0Y0}) and our symmetry assumptions on
$|A_1|^2, |A_2|^2$ and $A_1\bar{A}_2+\bar{A}_1A_2$ (see
(\ref{assum_A})), the above integral can be simplified to be
\begin{eqnarray}
I_1(x_0,y_0) =W_{1,1} \sin \left(2\pi x_0/L\right) +W_{1,2} \sin
\left(2\pi y_0/L\right),
\end{eqnarray}
where
\begin{equation}
W_{1,1} \equiv
\epsilon^2\int_{-\infty}^{+\infty}\int_{-\infty}^{+\infty}
\left\{c_{1,0}^{(1)} |A_1(X,Y)|^2+c_{1,0}^{(2)} |A_2(X,Y)|^2 +
c_{1,0}^{(3)}\left[A_1(X,Y)\bar{A}_2(X,Y)+\bar{A}_1(X,Y)A_2(X,Y)\right]\right\}
\cos \frac{2\pi x}{L} dxdy,
\end{equation}
and
\begin{equation}
W_{1,2}\equiv
\epsilon^2\int_{-\infty}^{+\infty}\int_{-\infty}^{+\infty}d_{1,0}^{(3)}
\left[A_1(X,Y)\bar{A}_2(X,Y)+\bar{A}_1(X,Y)A_2(X,Y)\right]\cos
\frac{2\pi y}{L} dxdy.
\end{equation}
Notice that both integrals $W_{1,2}$ and $W_{1,2}$ are exponentially
small in $\epsilon$, thus the constraint (\ref{constraint1}) is
exponentially small, hence it can not be captured by power-series
perturbation expansions (\ref{uexpand}). Repeating similar
calculations for the integral of $I_2(x_0,y_0)$ in Eq.
(\ref{constraint2}), we can get (to the leading order)
\begin{eqnarray}
I_2(x_0, y_0) =W_{2,1} \sin (2\pi x_0/L) +W_{2,2} \sin (2\pi y_0/L),
\end{eqnarray}
where expressions for $W_{2,1}$ and $W_{2,2}$ are similar to those
for $W_{1,1}$ and $W_{1,2}$ above. Then in order for the two
constraints (\ref{constraint1}) and (\ref{constraint2}) to hold, we
must have
\begin{equation}
\sin (2\pi x_0/L) =\sin (2\pi y_0/L)=0.
\end{equation}
Thus, the envelope solution $(A_1, A_2)$ can only be centered at
four locations:
\begin{equation} \label{locations}
(x_0, y_0) =(0, 0), \; (0, \frac{L}{2}),  \; (\frac{L}{2}, 0), \;
(\frac{L}{2}, \frac{L}{2}).
\end{equation}
Note that due to the $x$ and $y$ symmetry of the lattice, envelope
solutions centered at the second and third locations as above are
topologically the same (except an interchange of $x$ and $y$ axes).
Other possible locations of $(x_0, y_0)$, such as $(L, L)$, are
equivalent to one of the four locations above due to periodicity of
the lattice potential.

\section{Solutions of envelope equations}
\label{envelope_section}

Envelope equations (\ref{A})-(\ref{B}) and (\ref{A_single}) are one
of the key results of this article. They have important
consequences. First, they show that solitary waves are possible only
when $\eta D_1 <0$ and $\eta D_2 <0$. This simply means that $\mu$
must lie in the bandgap of the linear system (see Eq.
(\ref{muexpand})). Second, these equations show that solitary waves
exist only when the dispersion coefficients $D_1, D_2$ and the
nonlinearity coefficient $\sigma$ are of the same sign. For
instance, at points A, C and E in Fig. \ref{2Ddispersion} where
$D_1>0$ and $D_2>0$, solitary waves exist only when $\sigma>0$,
i.e., for focusing nonlinearity, but not for defocusing nonlinearity
($\sigma<0$). The situation is opposite at points B and D. This
result is consistent with our physical intuitions as well as the 2D
experimental observations in \cite{Kivshar_OE}.

Below, we consider soliton solutions of envelope equations
(\ref{A_single}) and (\ref{A})-(\ref{B}). The scalar equation
(\ref{A_single}) is the familiar 2D NLS equation, and it admits the
following types of solitary wave solutions:
\renewcommand{\labelenumi}{(\theenumi)}
\begin{enumerate}
\item $A_1 = f(R)$, which is real and radially symmetric; here $(R, \Theta)$ is the polar
coordinates of $(X, Y)$;
\item $A_1=f(R) e^{in\Theta}$, where $n$ is an integer;
this is a vortex-ring solution with charge $n$.
\end{enumerate}
The coupled envelope equations (\ref{A})-(\ref{B}) admit a wider
array of solutions. A complete classification of their solutions is
beyond the scope of the present article. Below we only list a few
types of their solutions. If $\gamma=0$, these equations admit the
following three simple solution reductions:
\renewcommand{\theenumi}{\roman{enumi}}
\begin{enumerate}
\item $A_1\ne 0$, $A_2=0$; or $A_1=0, A_2\ne 0$. In this case, the solution is a single Bloch-wave envelope
solution. Notice that $A_1$ (or $A_2$) satisfies a 2D NLS equation
with different dispersion coefficients $D_1$ and $D_2$ along the $X$
and $Y$ directions, thus these solutions are simply those described
above (charge-free or vortex solutions) except being stretched along
the $X$ or $Y$ direction, hence becoming ellipse-shaped.
\item $A_1, A_2 \in {\mathbb R}$, where $\mathbb R$ is the set of
real numbers. In this case, the solution is a real-valued vector
envelope state. Note that if $(A_1, A_2) \in {\mathbb R}$ is a
solution, so are $(-A_1, A_2), (A_1, -A_2), (-A_1, -A_2)$, and
$(iA_1, iA_2)$. All these solutions are equivalent to each other and
lead to the equivalent solitary waves in the original system
(\ref{u}).
\item $A_1 \in {\mathbb R}, A_2 \in i{\mathbb R}$. In this case, the solution
is a complex-valued vector envelope state. Note that the other
solution of $A_1\in i{\mathbb R}, A_2 \in {\mathbb R}$ is equivalent
to this $A_1 \in {\mathbb R}, A_2 \in i{\mathbb R}$ solution.
\end{enumerate}
If $\gamma\ne 0$, however, the reductions are quite different. For
instance, the first and third reductions of case $\gamma=0$ no
longer hold. In this case, the reduction of $A_1, A_2 \in {\mathbb
R}$ is allowed. Under this reduction, there are two subcases,
$A_1>0$, $A_2>0$, and $A_1>0, A_2<0$, which are \emph{not}
equivalent to each other. They lead to different solitary waves in
Eq. (\ref{u}) (see the end of this section).

It is note-worthy that at band edges where $\gamma\ne 0$, the single
Bloch-wave envelope reductions of $A_1\ne 0$, $A_2=0$ and $A_1=0,
A_2\ne 0$ are not possible. Physically, this is due to a resonance
between the two Bloch modes, which prevents the existence of a
single Bloch mode envelope solution. For instance, at point $E$ of
Fig. \ref{2Ddispersion} where $\gamma\ne 0$ (see Eq. (\ref{Edata})),
the two Bloch modes are both located at the same $\Gamma$ point in
the first Brillouin zone (see Sec. \ref{blochbands} and Fig.
\ref{Bloch_CD}(E)). These two Bloch modes are thus in resonance,
which makes $\gamma \ne 0$. At points where $\gamma= 0$ (such as
points $C$ and $D$ in Fig. \ref{2Ddispersion}), the two Bloch
solutions are located at different symmetry points of the Brillouin
zone and are not in resonance, thus single Bloch-wave reductions of
$A_1\ne 0$, $A_2=0$ and $A_1=0, A_2\ne 0$ are possible there.

To illustrate various envelope-soliton solutions admitted by Eqs.
(\ref{A_single}) and (\ref{A})-(\ref{B}), we consider points $A, B,
C, D$ and $E$ of Fig. \ref{2Ddispersion} in detail below. At each of
these five points, we determine solutions of the underlying envelope
equations. For each envelope solution, we also display the
corresponding leading-order analytical solution $u_0(x, y)$. Since
envelope solitons can have four different locations (see the
previous section), which will lead to four different solutions
$u_0(x, y)$, for simplicity, we will only display the one where the
envelope is centered at $(x_0, y_0)=(0, 0)$ below.

First we consider point A. At this point, a single Bloch mode exists
and has been shown in Fig. \ref{Bloch_CD}(A). This Bloch wave is
located at $\Gamma$ point of the first Brillouin zone, is
$\pi$-periodic along both $x$ and $y$ directions, and is all
positive. The envelope equation at this point is given by
(\ref{A_single}) with the coefficients $D_1$ and $\alpha_0$ given in
Eq. (\ref{Acoef}). Since $D_1>0$, solitary waves will bifurcate into
the semi-infinite bandgap ($\eta <0$) under {\it focusing}
nonlinearity ($\sigma=1$). If we take the solution of
(\ref{A_single}) to be $A_1=f(R)$ where $f(R)>0$ is the ground-state
envelope solution, the corresponding leading-order analytical
solution $u_0(x, y)$ is a nodeless solitary wave. Such solutions
have been reported before
\cite{Segev03,YangMuss,Christdoulides03,ChenMartin}. Other envelope
solutions of (\ref{A_single}) such as $A_1=f(R)$ where $f(R)$ is a
higher-mode solution (with nodes) or vortex-ring solutions $A_1=f(R)
e^{in\Theta}$ could lead to other types of solitary wave structures.

Next, we consider point B. At this point, a single Bloch mode exists
and has been shown in Fig. \ref{Bloch_CD}(B). This Bloch mode is
located at $M$ point of the first Brillouin zone, is $2\pi$-periodic
along both $x$ and $y$ directions, and its adjacent peaks are out of
phase. The envelope equation at this point is given by
(\ref{A_single}) with the coefficients $D_1$ and $\alpha_0$ given in
Eq. (\ref{Bcoef}). Since $D_1<0$, solitary waves will bifurcate into
the first bandgap ($\eta >0$) under {\it defocusing} nonlinearity
($\sigma=-1$). If we take the solution of (\ref{A_single}) to be the
ground-state envelope solution, the corresponding analytical
solution $u_0(x, y)$ is a gap soliton with nodes. Such solutions
have been reported in
\cite{Segev03,Christdoulides03,Kivshar_Opt_Exp}.

Envelope solutions at points C, D and E are richer and more
interesting. Two of them, which are the so-called reduced-symmetry
solitons \cite{Kivshar_C_onemode} and gap vortex solitons
\cite{Segev_highervortex}, have been reported before. But many other
solutions at these points are novel and have not been considered
yet. They include solutions which we call dipole-array solitons,
dipole-cell solitons, vortex-cell solitons, etc. Envelope solutions
and the corresponding leading-order analytical solutions $u_0(x, y)$
at these three points will be described in the next three
subsections respectively.

\subsection{Envelope solutions at point C}
At point C, two Bloch modes exist. One of them is shown in Fig.
\ref{Bloch_CD}(C), while the other is a $90^\circ$-rotation of Fig.
\ref{Bloch_CD}(C). These Bloch modes are located at the $X$ and $X'$
points of the first Brillouin zone. They are $\pi$-periodic along
one spatial direction, and $2\pi$-periodic along the other spatial
direction. Envelope equations at this point are given by
(\ref{A})-(\ref{B}), with the coefficients $D_1, D_2, \alpha, \beta$
given in Eq. (\ref{Cdata}), and $\gamma=0$. Since $D_1, D_2>0$ here,
solitary waves will bifurcate into the first bandgap ($\eta <0$)
under {\it focusing} nonlinearity ($\sigma=1$). Since $\gamma=0$, we
have three solution reductions (see above). Under these reductions,
we consider the following subclasses of solutions.

\renewcommand{\theenumi}{\roman{enumi}}
\begin{enumerate}
\item $A_1>0$, $A_2=0$. In this case,
the envelope solution is an ellipse as shown in Fig.
\ref{soliton_C_anal0}(a). This ellipse stretches along the $Y$
direction. When the Bloch wave (see Fig. \ref{Bloch_CD}(C)) is
modulated by this envelope function, the resulting leading-order
analytical solution $u_0(x, y)$ is plotted in Fig.
\ref{soliton_C_anal0}(b) (with $\e$ taken as 0.2). This solution
contains only a single Bloch mode (since $A_2=0$), thus we can call
it a {\it single-Bloch-mode soliton}. This soliton is narrower along
the $x$ direction, and broader along the $y$ direction, thus it is
expected to be more mobile along the $y$ direction and less so along
the $x$ direction. This type of solution has been observed in
\cite{Kivshar_C_onemode} for a different (saturable) nonlinearity
(see Sec. \ref{vortex_section} (A) for more details). In that paper,
these solutions were called reduced-symmetry solitons. One of their
potential applications in optical routing and switching has been
described in \cite{Christodoulides_application}.

\begin{figure}
\center
\includegraphics[width=0.4\textwidth]{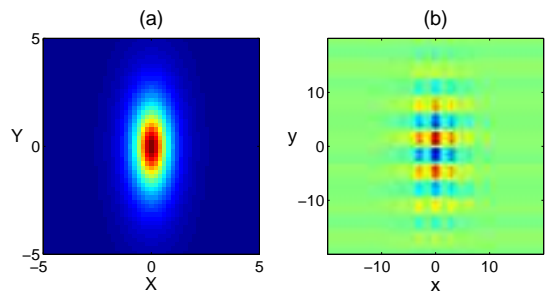}
\caption{(color online) (a) Envelope solution $A_1(X, Y)$ at point C
with $A_2=0$; (b) the corresponding analytical single-Bloch-mode
soliton $u_0(x, y)$ with $\e=0.2$. \label{soliton_C_anal0}}
\end{figure}

\item $A_1>0$, $A_2>0$. In this case, the envelope solutions are both real and positive,
and they are shown in Fig. \ref{soliton_C_anal}(a, b). They are both
ellipse-shaped but stretched along opposite directions. The
corresponding leading-order analytical solution $u_0(x, y)$ is
plotted in Fig. \ref{soliton_C_anal}(c, d) (with $\e$ taken as 0.2).
The central region of this solitary wave is a dipole array aligned
along the two diagonal directions, thus we will call this solitary
wave a {\it dipole-array gap soliton}. The outer region of this
soliton is aligned along the horizontal ($x$) and vertical ($y$)
directions though. Note that this dipole-array soliton is quite
different from dipole solitons reported before (see
\cite{Neshev,YangStudies} for instance): previous dipole solitons
reside in the semi-infinite bandgap, and their peaks are at lattice
sites; but the present dipole-array soliton resides in the first
bandgap, and its peaks are off lattice sites. This dipole-array gap
soliton arises due to a superposition of two modulated Bloch modes
(see Eq. (\ref{u0})), and has never been reported before.

\item $A_1>0$, $A_2=i\hat{A}_2$, $\hat{A}_2>0$. In this case, the
envelope of one Bloch wave is real, while that of the other Bloch
wave purely imaginary. In the literature, such two Bloch modes are
called to have $\pi/2$ phase delay
\cite{Segev_highervortex,MakazyukPRL}. Envelope functions $A_1$ and
$\hat{A}_2$ look very similar to $A_1$ and $A_2$ of Fig.
\ref{soliton_C_anal}(a, b). Indeed, it is easy to see that $A_1$ and
$\hat{A}_2$ satisfy the same equations as $A_1$ and $A_2$ of the
previous reduction (ii) except that the $\beta$ coefficient is
slightly different. The leading-order analytical solution $u_0(x,
y)$ for these envelope solutions is displayed in Fig.
\ref{soliton_C_anal}(e, f). This soliton looks quite different from
the previous one in Fig. \ref{soliton_C_anal}(c, d). The most
significant difference is that, when winding around each lattice
center (i.e. points $x=m\pi, y=n\pi$ with $m, n$ being integers),
the phase of the present soliton increases or decreases by $2\pi$.
In other words, the solution around each lattice center has a vortex
structure. Thus we call this solution a {\it vortex-array gap
soliton}. This vortex-array soliton is qualitatively the same as the
gap vortex soliton reported in \cite{Segev_highervortex} for a
different nonlinearity (see Sec. \ref{vortex_section}(A)). However,
it is quite different from the other vortex solitons residing inside
the semi-infinite bandgap, as has been reported in
\cite{YangMuss,Malomed,Chen_vortex,Segev_vortex} before.

For a vortex-type soliton, its angular momentum is an important
quantity. This momentum is defined as
\begin{equation}
\Omega \equiv \int \mathbf{r}\times \mbox{Im}(\bar{u}\nabla u) \:
d\mathbf{r}=\mbox{Im} \int_{-\infty}^\infty \int_{-\infty}^\infty
\bar{u}(xu_y-yu_x) dxdy.
\end{equation}
Here $\mbox{Im}(\bar{u}\nabla u)$ is the linear momentum, and
$\mathbf{r}$ is the position vector in the $(x, y)$ plane. The spin
of the vortex is defined as
\begin{equation}
S=\frac{\Omega}{\int |u|^2 \: d\mathbf{r}}.
\end{equation}
In the absence of the periodic potential, the spin of a vortex
soliton takes an integer value equal to the phase winding number
(topological charge). In the present case with a periodic potential,
the spin of a vortex-array gap soliton does not take integer values
in general. Even though this soliton has a local vortex structure
around each lattice center, the topological charges of adjacent
lattice centers are opposite, thus the total spin of this
vortex-array soliton is finite, not infinite. In fact, substituting
the perturbation-series solution (\ref{uexpand}), (\ref{u0}), and
$A_1\in \mathbb{R}$, $A_2\in i\mathbb{R}$ into the $\Omega$ and $S$
formulas, we can easily show that both $\Omega$ and $S$ approach
zero as $\e \to 0$. Thus the spin of these vortex-array solitons is
quite small.

\begin{figure}
\center
\includegraphics[width=0.4\textwidth]{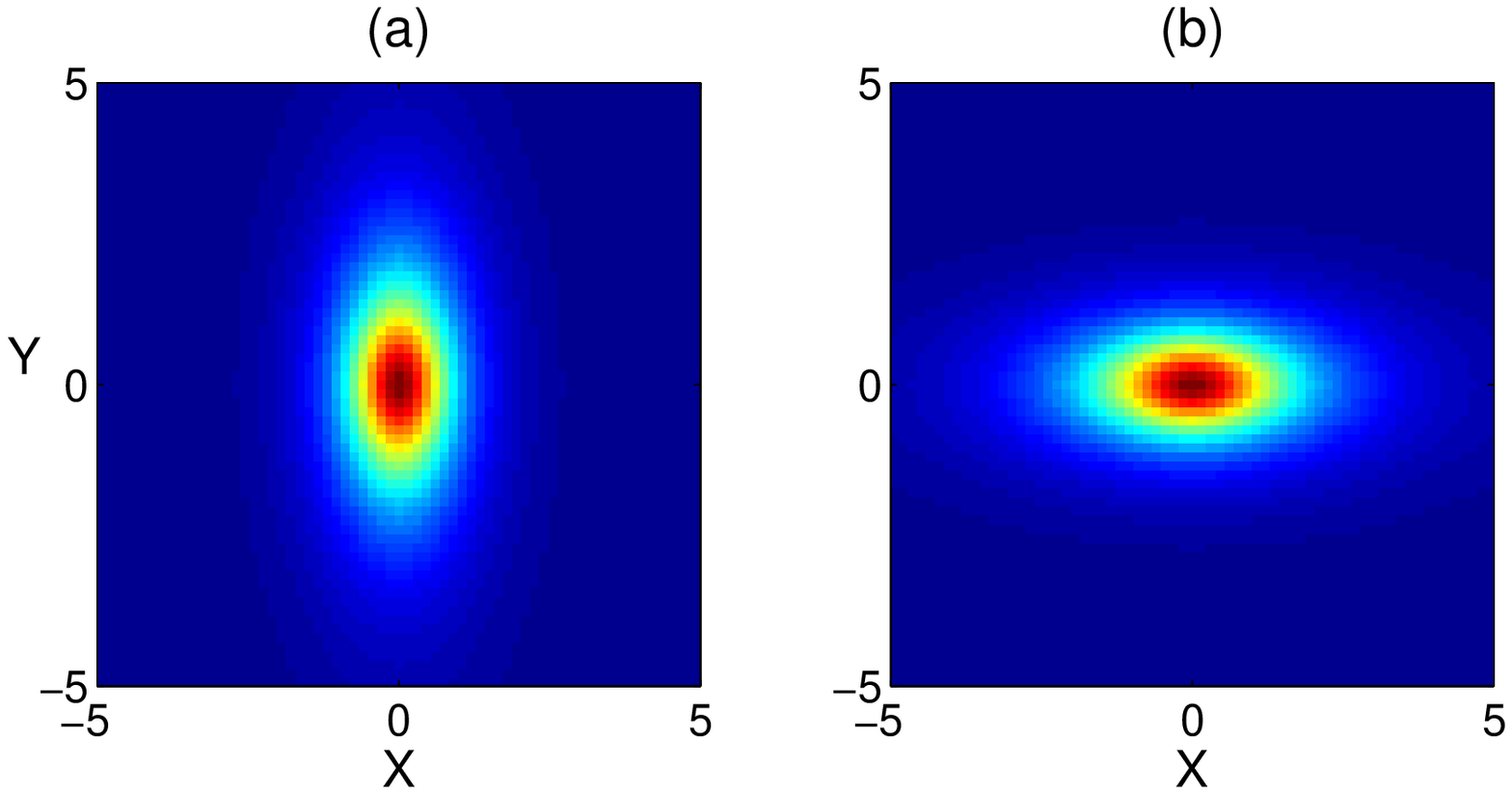}

\includegraphics[width=0.4\textwidth]{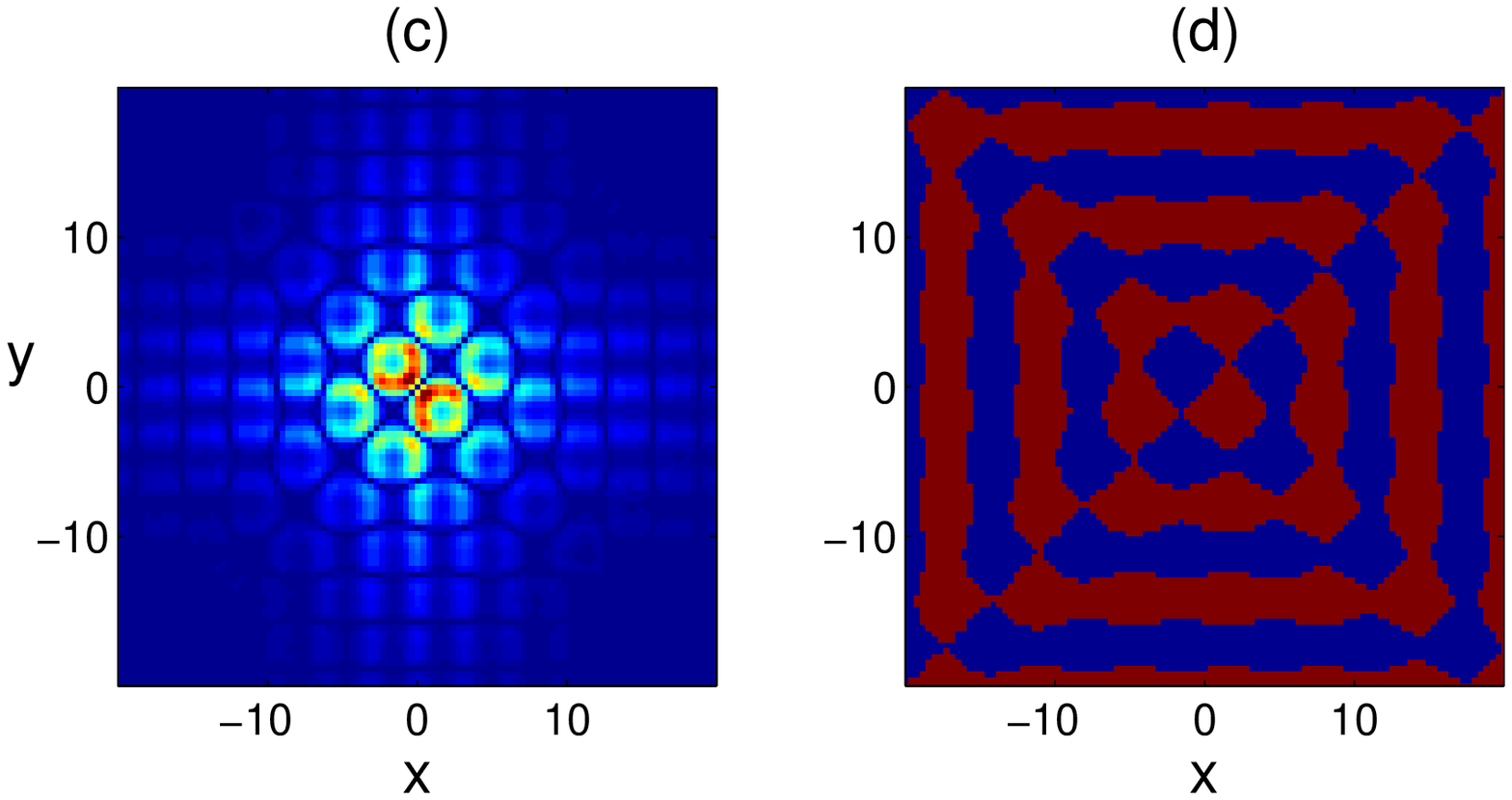}

\includegraphics[width=0.4\textwidth]{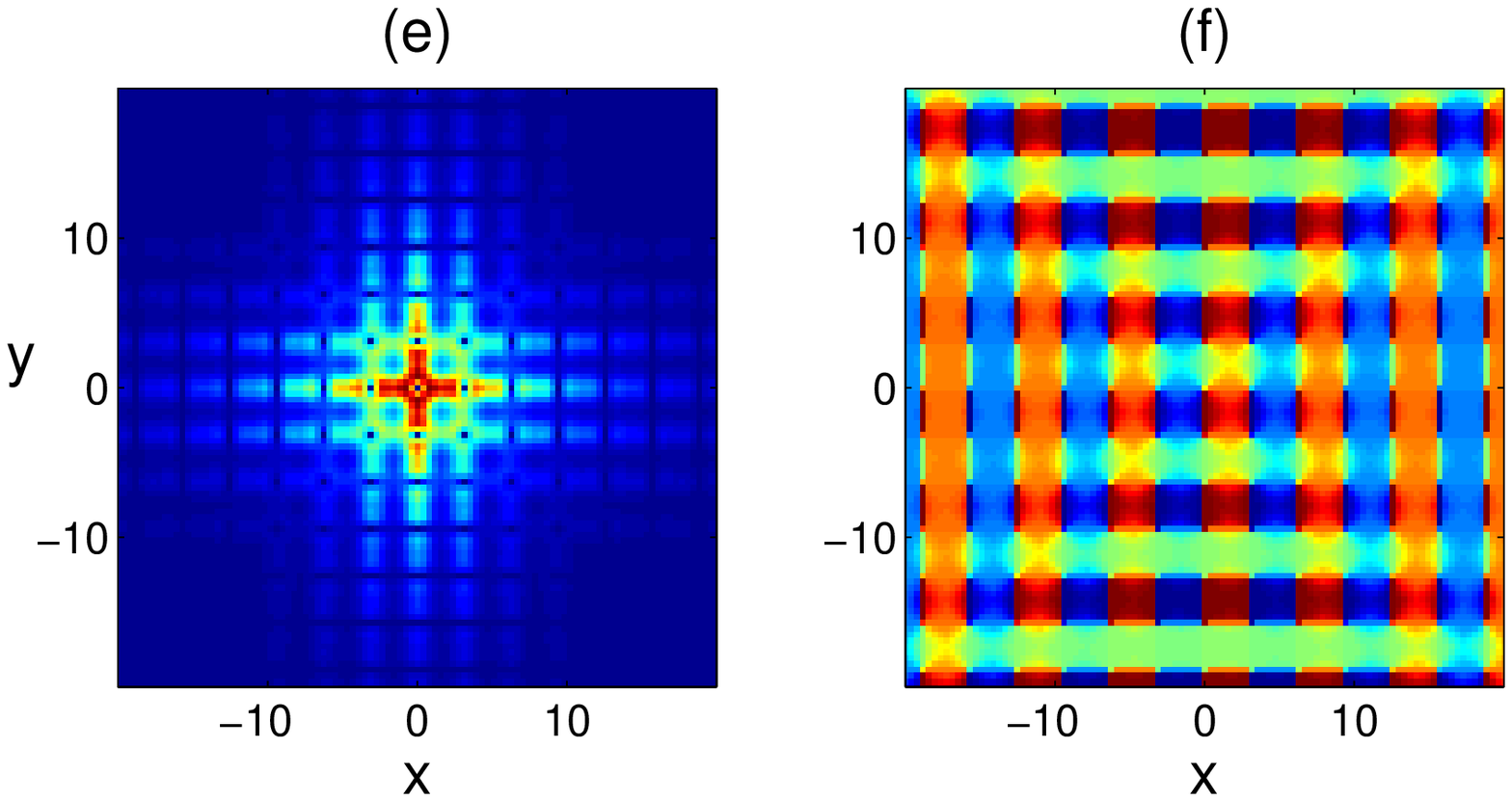}
\caption{(color online) (a, b) Envelope solutions $A_1(X, Y)>0$
(left) and $A_2(X, Y)>0$ (right) at point C;  (c, d) the
corresponding analytical dipole-array soliton $u_0(x, y)$ with
$\e=0.2$: the left is the amplitude and right the phase;  (e, f) the
analytical vortex-array soliton $u_0(x, y)$ at point C in the case
(iii) with $\e=0.2$: the left is the amplitude and right the phase.
\label{soliton_C_anal}}
\end{figure}

\end{enumerate}

\subsection{Envelope solutions at point D}
Point D admits two Bloch waves, one shown in Fig. \ref{Bloch_CD}(D),
and the other one being a $90^\circ$-rotation of Fig.
\ref{Bloch_CD}(D). These Bloch modes are located at the $X$ and $X'$
points of the first Brillouin zone.  Envelope equations at this
point are given by (\ref{A})-(\ref{B}), with the coefficients $D_1,
D_2, \alpha, \beta$ given in Eq. (\ref{Ddata}), and $\gamma=0$.
Since $D_1, D_2<0$ now, solitary waves will bifurcate into the
second bandgap ($\eta >0$) under {\it defocusing} nonlinearity
($\sigma=-1$). Like point C, three solution reductions are admitted.
Under these reductions, we consider the following subclasses of
solutions.

\begin{enumerate}
\item $A_1>0$, $A_2=0$. In this case, the envelope solution is an
ellipse as shown in Fig. \ref{soliton_D_anal0}(a), which is thinner
than Fig. \ref{soliton_C_anal0}(a) at point C. When the Bloch wave
(see Fig. \ref{Bloch_CD}(D)) is modulated by this envelope function,
the resulting leading-order analytical solution $u_0(x, y)$ is
plotted in Fig. \ref{soliton_D_anal0}(b) (with $\e=0.2$). This
solution is also a single-Bloch-mode soliton since $A_2=0$. Due to
the thin envelope solution, this soliton is much broader in the $y$
direction than in the $x$ direction. Thus we can expect it to be
much more mobile along the $y$ direction than along the $x$
direction. Note that this solution is under defocusing nonlinearity,
while a similar-looking solution in Fig. \ref{soliton_C_anal0}(b)
(see also \cite{Kivshar_C_onemode}) was under focusing nonlinearity.

\begin{figure}
\center
\includegraphics[width=0.4\textwidth]{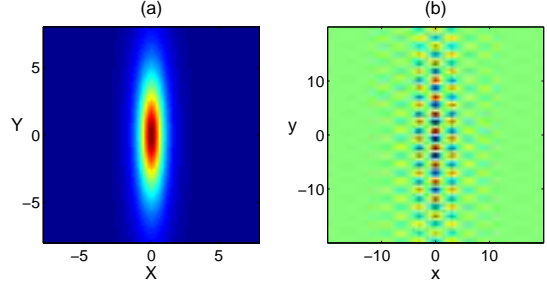}
\caption{(color online) (a) Envelope solution $A_1(X, Y)$ at point D
with $A_2=0$; (b) the corresponding analytical single-Bloch-mode
soliton $u_0(x, y)$ with $\e=0.2$. \label{soliton_D_anal0}}
\end{figure}

\item $A_1>0$, $A_2>0$. In this case, the envelope solutions are both real and positive,
and are shown in Fig. \ref{soliton_D_anal}(a, b). They are both
ellipse-shaped stretching along opposite directions. The
corresponding leading-order analytical solution $u_0(x, y)$ is
plotted in Fig. \ref{soliton_D_anal}(c, d) (with $\e=0.2$). This
solution looks quite different from its counterpart --- the
dipole-array soliton of Fig. \ref{soliton_C_anal}(c, d) at point C.
Its main feature is that the solution inside each lattice site is a
dipole cell. Its difference from the dipole-array soliton at point C
is that, here the two humps of each dipole are completely confined
inside each individual lattice, while the two humps of each dipole
in Fig. \ref{soliton_C_anal}(c, d) spread to neighboring lattice
sites. We will call this soliton in Fig. \ref{soliton_D_anal}(c, d)
a {\it dipole-cell gap soliton}. Solutions of this kind have not
been reported before. Notice that in the central region of the
soliton, dipole cells are aligned along diagonal directions, but in
the outer region, dipole cells are dominated and aligned along the
horizontal and vertical directions.

\begin{figure}
\center
\includegraphics[width=0.4\textwidth]{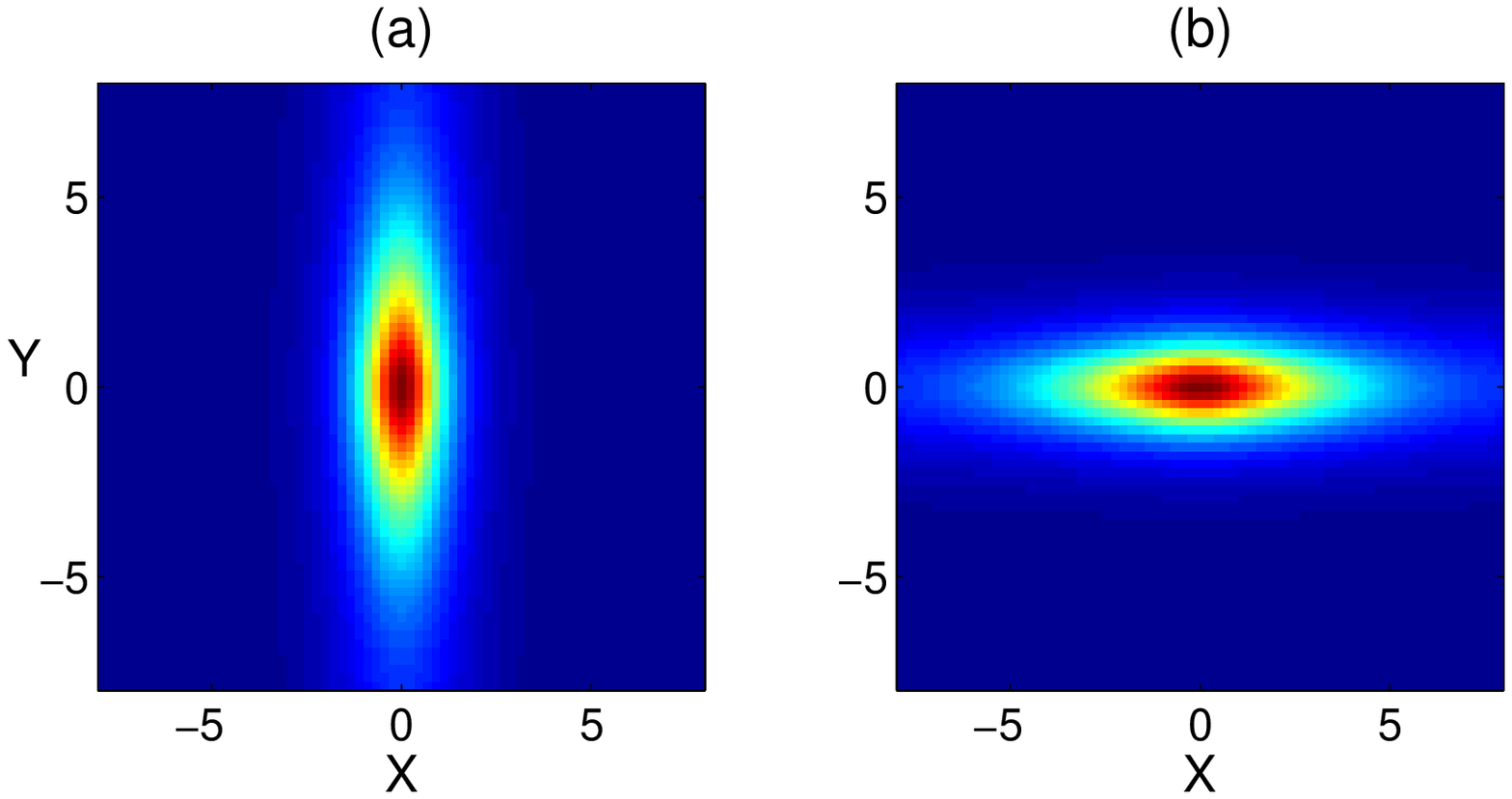}

\includegraphics[width=0.4\textwidth]{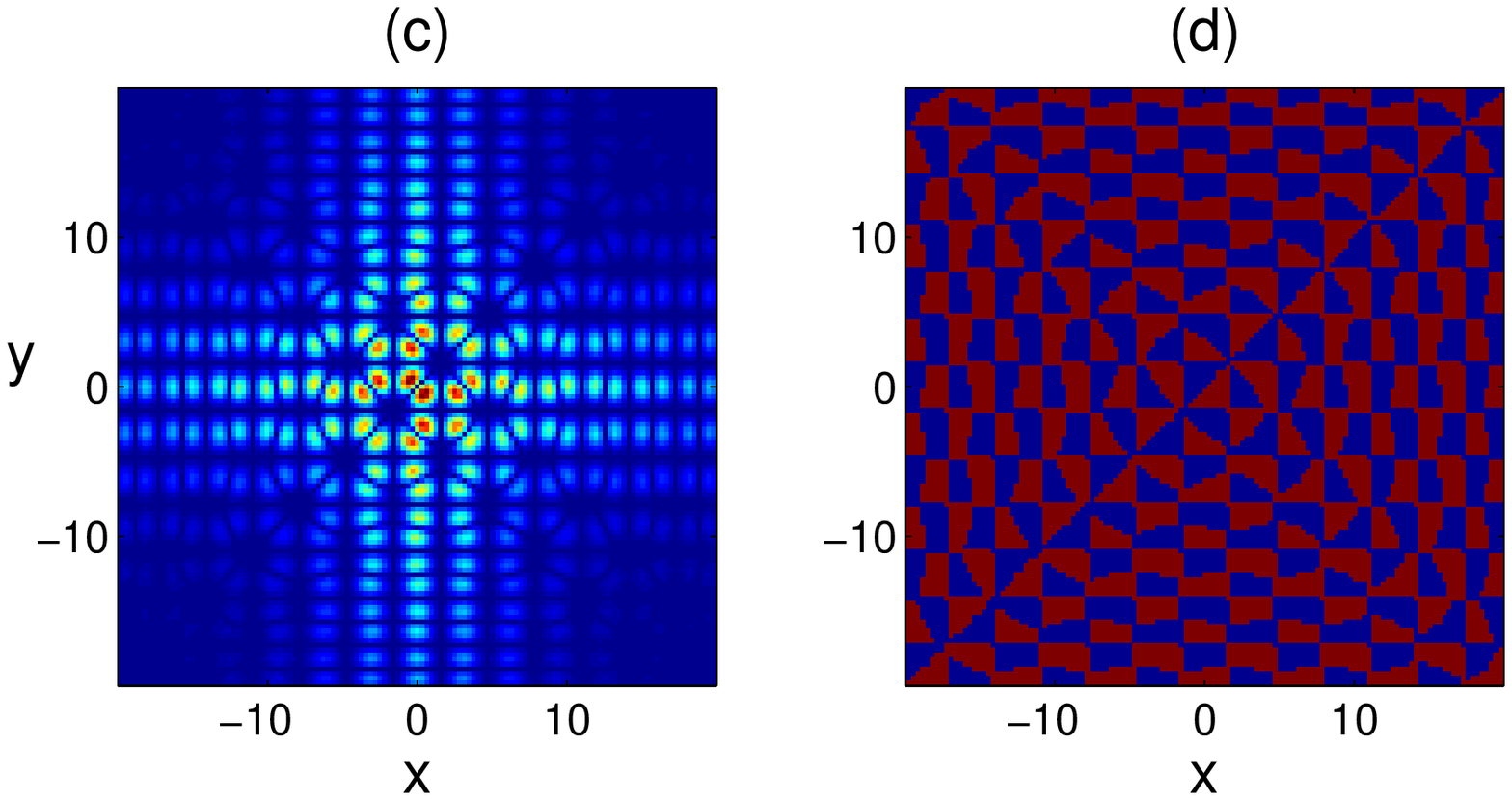}

\includegraphics[width=0.4\textwidth]{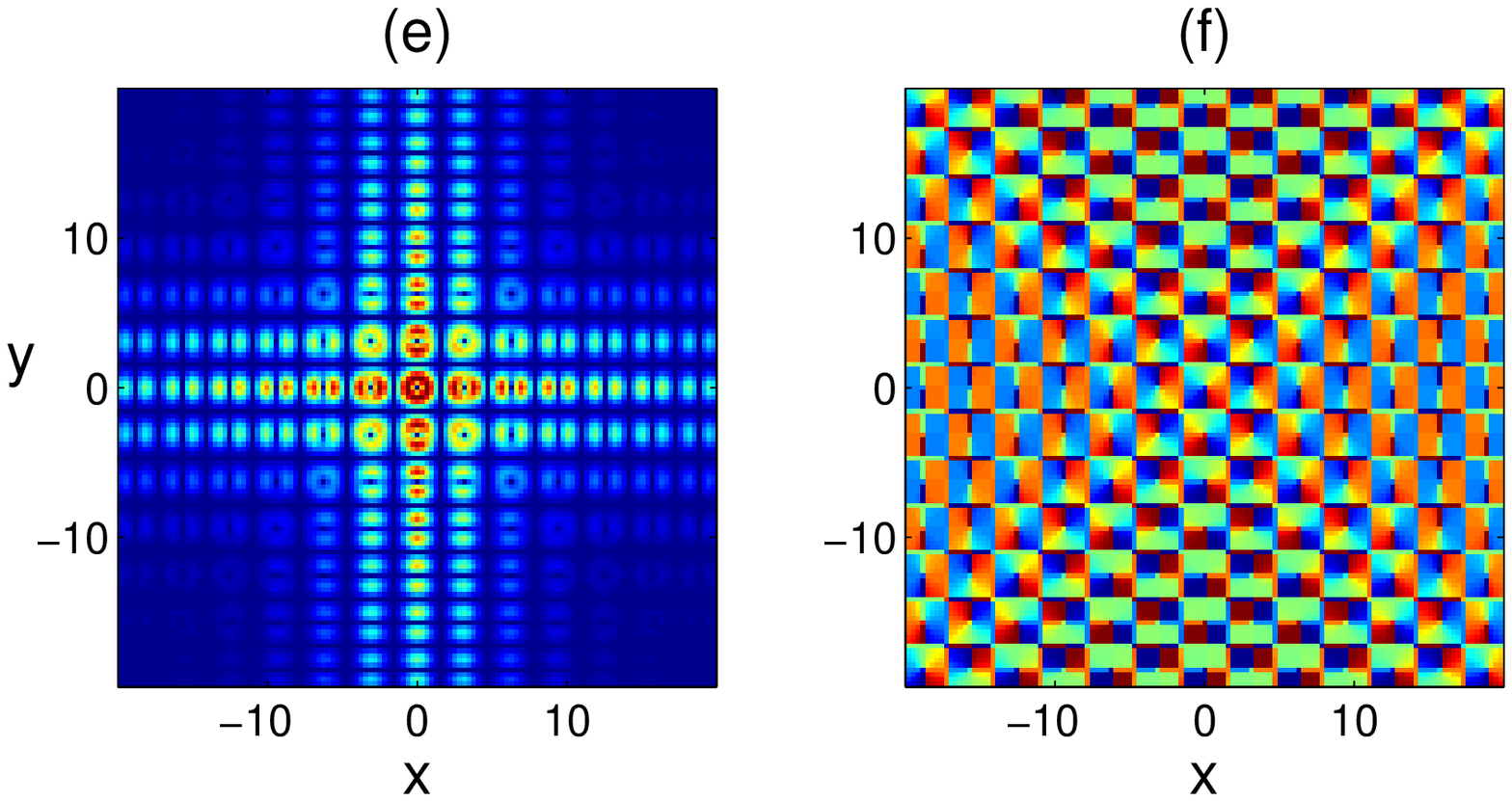}
\caption{(color online) (a, b) Envelope solutions $A_1(X, Y)>0$
(left) and $A_2(X, Y)>0$ (left) at point D;  (c, d) the
corresponding analytical dipole-cell soliton $u_0(x, y)$ with
$\e=0.2$: the left is the amplitude and right the phase; (e, f) the
analytical vortex-cell soliton $u_0(x, y)$ at point D in the case
(iii) with $\e=0.2$: the left is the amplitude and right the phase.
\label{soliton_D_anal}}
\end{figure}

\item $A_1>0$, $A_2=i\hat{A}_2$, $\hat{A}_2>0$. In this case, the two Bloch modes
have $\pi/2$ phase delay, and envelope functions $(A_1, \hat{A}_2)$
are similar to $(A_1, A_2)$ of Fig. \ref{soliton_D_anal}(a, b). The
leading-order analytical solution $u_0(x, y)$ with these envelope
solutions is displayed in Fig. \ref{soliton_D_anal}(e, f). The most
significant feature of this solution is that, in the central region,
the solution inside each lattice site is a vortex cell (with charge
1 or $-1$). Thus we call it a {\it vortex-cell gap soliton}. Charges
of adjacent cells are opposite, hence the total angular momentum and
spin of this soliton are both finite. When $\e \to 0$, the angular
momentum and spin approach zero. Notice that each vortex cell here
is completely isolated and confined inside each individual lattice,
which is quite different from the vortex-array soliton in Fig.
\ref{soliton_C_anal}(e, f), where neighboring vortices are connected
together and do not have clear-cut boundaries between them. In
addition, this vortex-cell soliton is under defocusing nonlinearity,
while the vortex-array soliton of Fig. \ref{soliton_C_anal}(e, f)
was under focusing nonlinearity (see also
\cite{Segev_highervortex}). Furthermore, this vortex-cell soliton
resides in the \emph{second} bandgap, while the vortex-array soliton
of Fig. \ref{soliton_C_anal}(e, f) resides in the first bandgap. Gap
vortex solitons under defocusing nonlinearity as reported in
\cite{Kivshar_Opt_Exp,Malomed_gap_vortex} reside also inside the
first bandgap; they are the counterparts of vortex solitons in the
semi-infinite bandgap as reported in \cite{YangMuss, Malomed}, and
are fundamentally different from the vortex-cell solitons bifurcated
from point D here. Although these vortex-cell gap solitons have
never been studied before, they do resemble the linear vortex-cell
defect modes in photonic lattices as reported in \cite{MakazyukPRL}.

\end{enumerate}

\subsection{Envelope solutions at point E}

Point E also admits two Bloch modes, one shown in Fig.
\ref{Bloch_CD}(E), and the other being a $90^\circ$-rotation of Fig.
\ref{Bloch_CD}(E). These Bloch modes are both located at $\Gamma$
point of the first Brillouin zone, thus are $\pi$-periodic along
both spatial directions. Envelope equations at this point are given
by (\ref{A})-(\ref{B}), with the coefficients $D_1, D_2, \alpha,
\beta, \gamma$ given in Eq. (\ref{Edata}). Since $D_1, D_2>0$,
solitary waves will bifurcate into the second bandgap ($\eta <0$)
under {\it focusing} nonlinearity ($\sigma=1$). Since $\gamma \ne
0$, the reduction of $A_1, A_2 \in {\mathbb R}$ is allowed, and we
consider the following two subclasses of solutions.

\begin{enumerate}
\item $A_1>0$, $A_2>0$. In this case, the envelope solutions are both real and positive,
and they are shown in Fig. \ref{soliton_E_anal}(a, b). They are both
very thin ellipses stretched along opposite directions. The
corresponding leading-order analytical solution $u_0(x, y)$ is
plotted in Fig. \ref{soliton_E_anal}(c) (with $\e$ taken as 0.2).
This solution has a pronounced cross-shape overall structure. In the
central region of the $(x, y)$ plane, the solution at each lattice
center is an almost-circular positive spike, which recedes to
negative backgrounds away from the lattice center. Thus we may call
this solution a spike-array gap soliton.

\item $A_1>0$, $A_2<0$. In this case, envelope functions $A_1$ and $|A_2|$ are
similar to those of Fig. \ref{soliton_E_anal}(a, b), thus not shown.
The corresponding leading-order analytical solution $u_0(x, y)$ is
plotted in Fig. \ref{soliton_E_anal}(d) (with $\e=0.2$). This
solution also has a pronounced cross shape. In its central region,
the solution resembles an array of quadrupoles, thus we can call
this solution a quadrupole-array gap soliton.

\begin{figure}
\center
\includegraphics[width=0.4\textwidth]{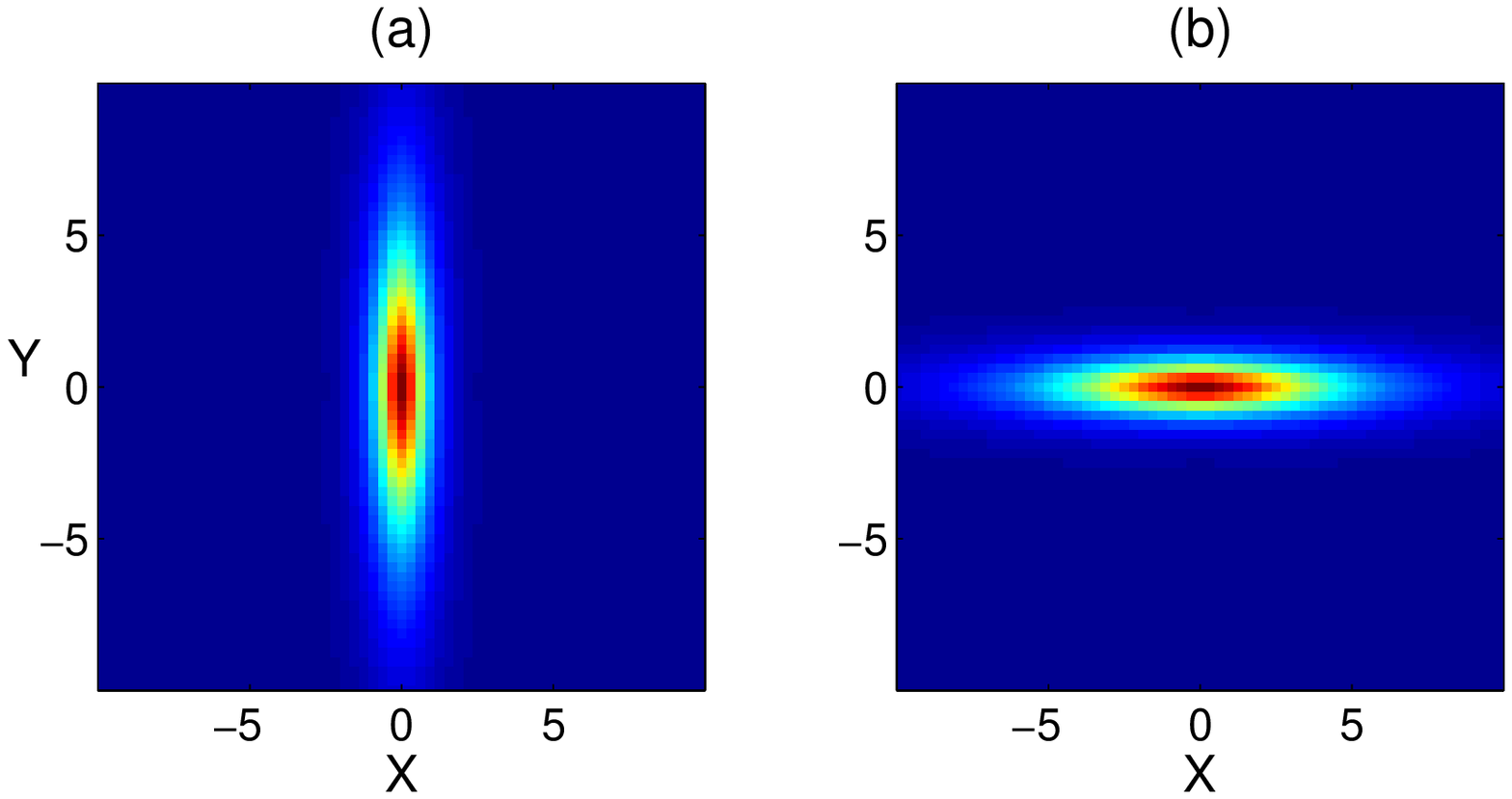}

\includegraphics[width=0.4\textwidth]{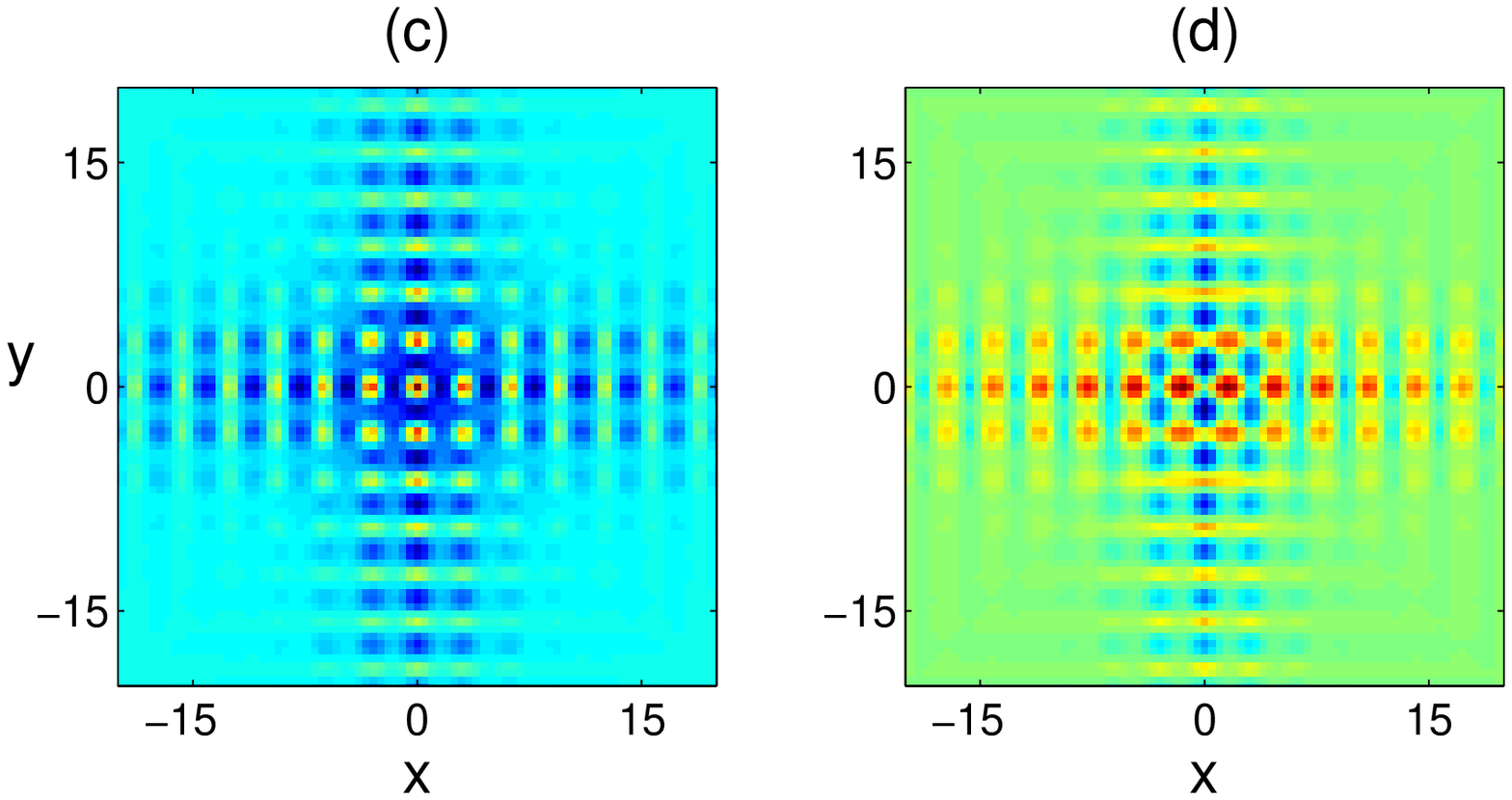}
\caption{(color online) (a, b) Envelope solutions $A_1(X, Y)>0$
(left) and $A_2(X, Y)>0$ (right) at point E;  (c) the corresponding
analytical spike-array soliton $u_0(x, y)$ with $\e=0.2$; (d) the
analytical quadrupole-array soliton $u_0(x, y)$ at point E in the
case (ii) with $\e=0.2$. \label{soliton_E_anal} }
\end{figure}

\end{enumerate}

\section{Families of solitary waves bifurcated from band edges}
\label{vortex_section}

The above multi-scale perturbation analysis predicts various types
of low-amplitude solitary waves when the propagation constant $\mu$
is near edges of Bloch bands. As the propagation constant moves away
from band edges, these solitary waves will become more localized,
and their amplitudes will increase. From an experimental point of
view, more localized solitary waves are often easier to observe. For
more localized solutions, the above perturbation analysis starts to
break down, and numerical methods need to be employed. In this
section, we compute whole families of solitary waves bifurcated from
edges of Bloch bands. The numerical method we will use is the
modified squared-operator iteration method developed in
\cite{YangLakoba}.

For illustration purpose, we determine families of solutions
bifurcated from points C, D and E of Fig. \ref{2Ddispersion}(b).
Leading-order analytical approximations on low-amplitude solutions
of these families have been presented in Figs. \ref{soliton_C_anal0}
to \ref{soliton_E_anal}. Note that solution families bifurcated from
points A and B have been reported before in
\cite{YangMuss,Christdoulides03,Kivshar_Opt_Exp}, thus they will not
be computed in this article.

\subsection{Solution families bifurcated from point C}
\begin{figure}
\center
\includegraphics[width=0.45\textwidth]{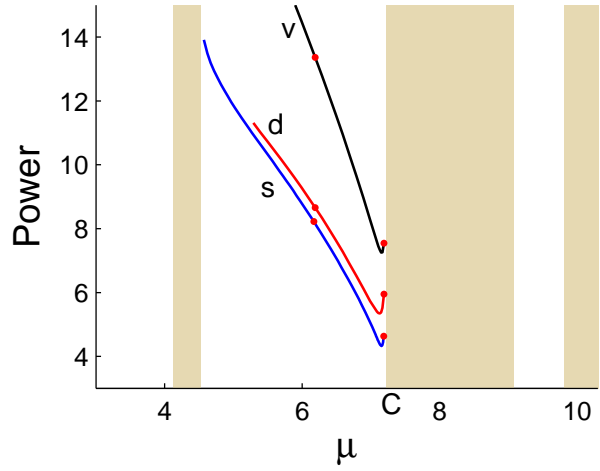}
\caption{(color online) Power diagrams of the three families of
solitary waves bifurcated from the edge point C (under focusing
nonlinearity). Bottom curve: single-Bloch-mode branch (marked by
letter 's'); middle curve: dipole-array branch (marked by letter
'd'); top curve: vortex-array branch (marked by letter 'v'). The
marked thick dot points are 0.04 and 1.04 below the band edge C.
\label{powerC}}
\end{figure}

\begin{figure}
\center
\includegraphics[width=0.4\textwidth]{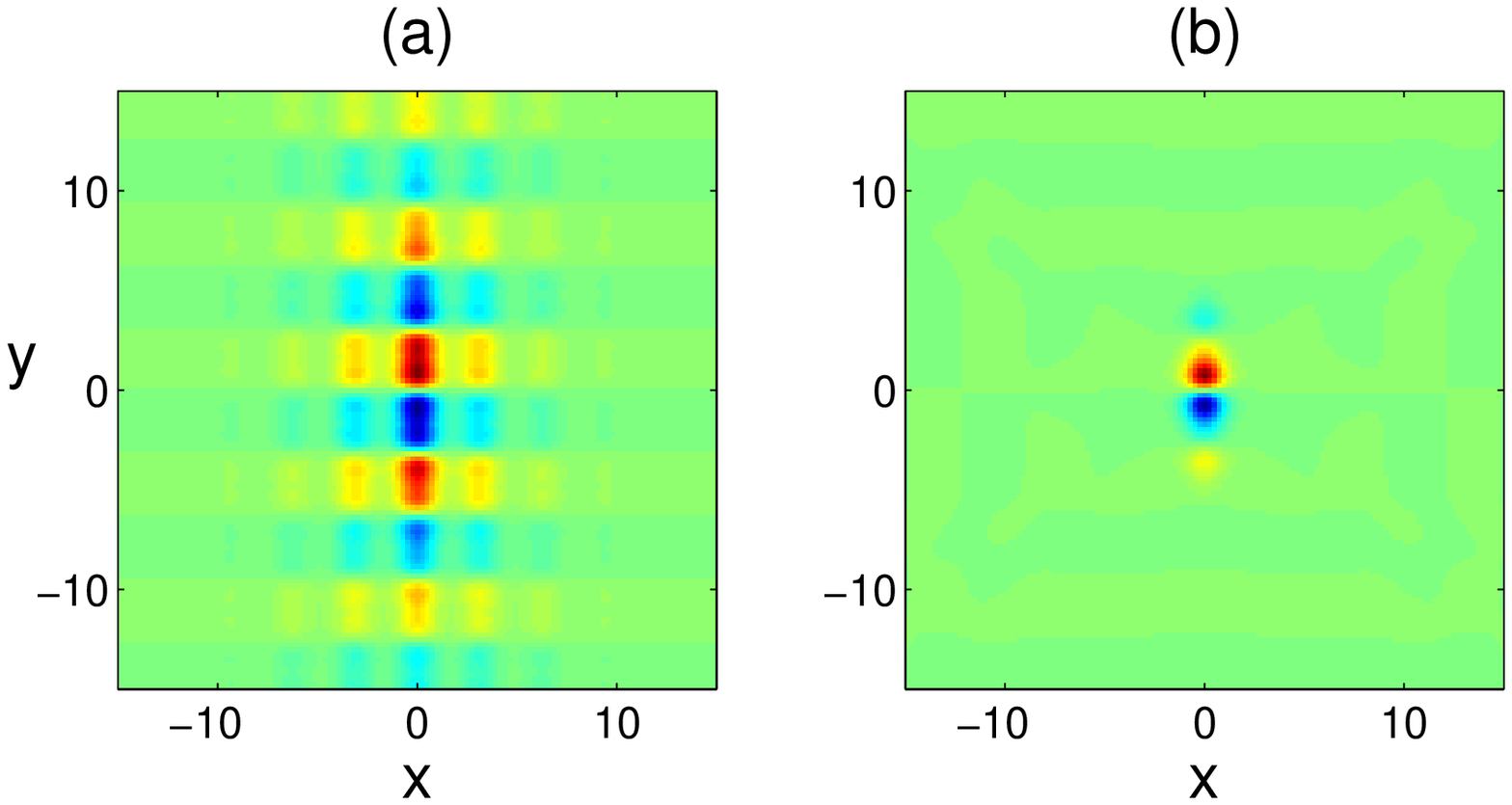}

\includegraphics[width=0.6\textwidth]{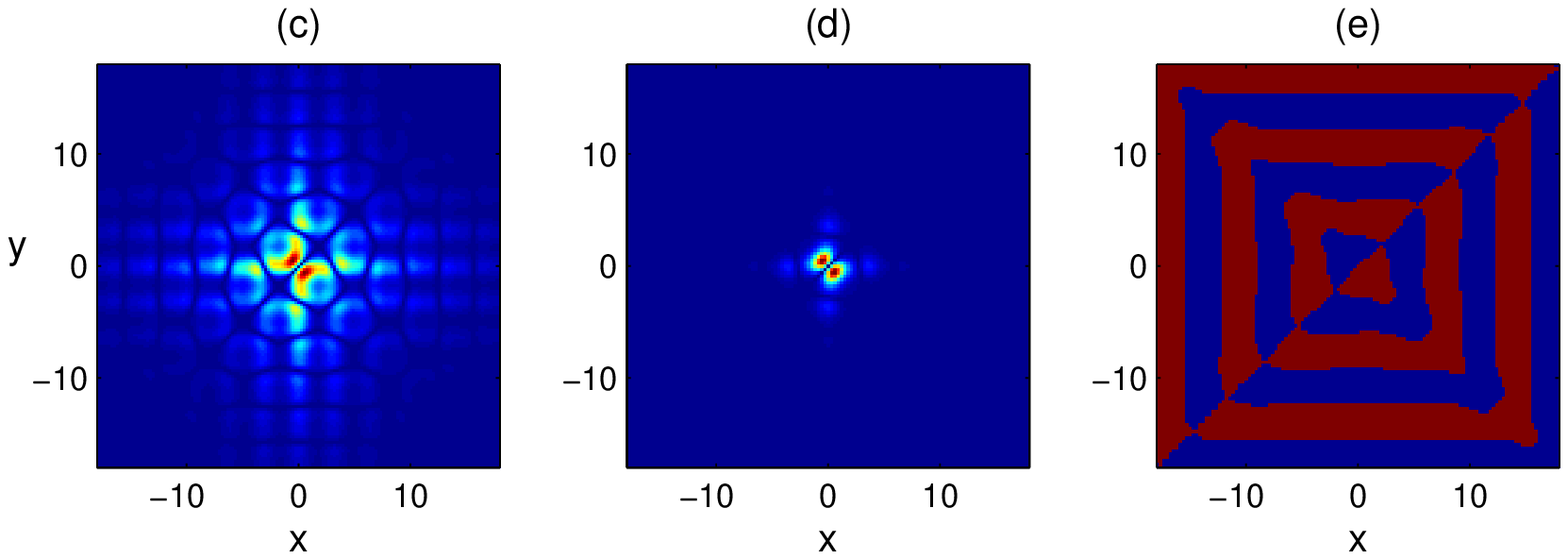}

\includegraphics[width=0.6\textwidth]{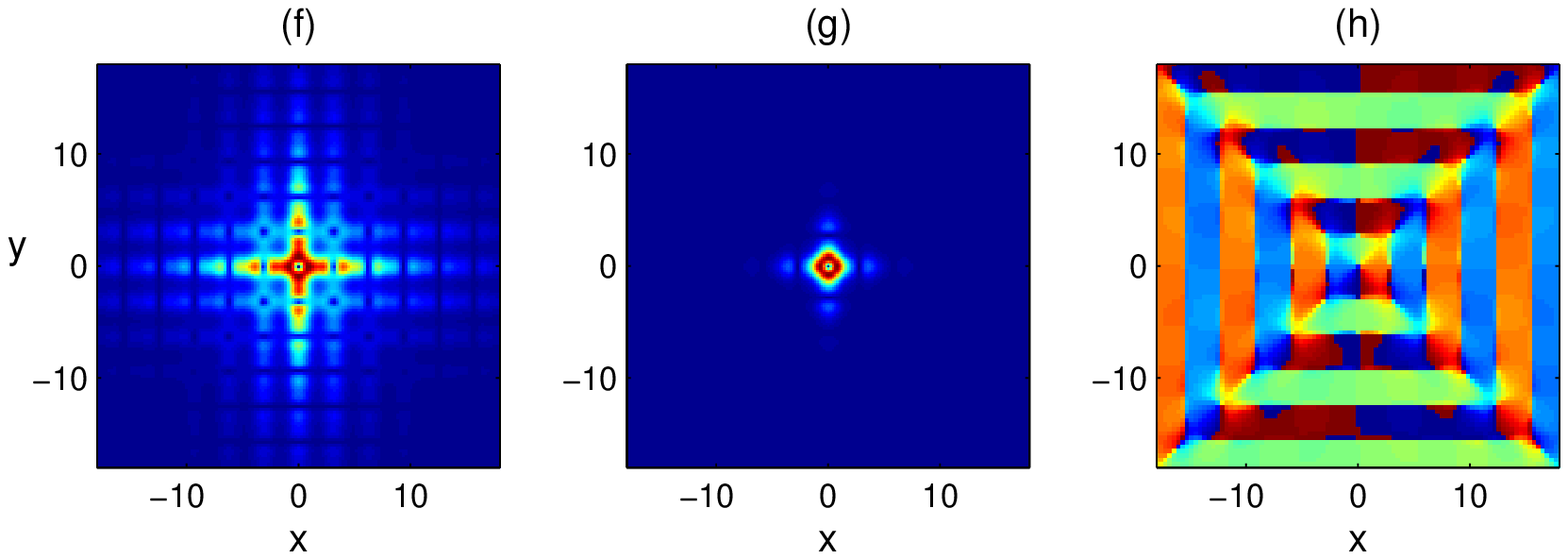}
\caption{(color online) Weakly and strongly localized solutions on
the three soliton branches bifurcated from the edge point C in Fig.
\ref{powerC}. Propagation constants of these solutions are
$\mu=6.189$ and 7.189, which are marked by thick dots in Fig.
\ref{powerC}. Top row: single Bloch-mode solitons; $u(x, y)$ is
shown; (a) $\mu=7.189$; (b) $\mu=6.189$. Middle and bottom rows:
dipole-array and vortex-array solitons respectively; (c, f)
amplitudes ($|u(x, y)|$) at $\mu=7.189$; (d, e) amplitude and phase
of the dipole-array soliton at $\mu=6.189$; (g, h) amplitude and
phase of the vortex-array soliton at $\mu=6.189$. \label{soliton_C}}
\end{figure}

At point C, three types of solitary waves have been predicted in the
previous section under focusing nonlinearity: single-Bloch-mode
solitons, dipole-array solitons, and vortex-array solitons, see
Figs. \ref{soliton_C_anal0}(b), \ref{soliton_C_anal}(c, d) and
\ref{soliton_C_anal} (e, f) respectively. They all exist in the
first bandgap, bifurcated from the edge point C of the second Bloch
band. Numerically, we have obtained the entire families of these
three types of solutions. Their power curves are displayed in Fig.
\ref{powerC}. Here the power is defined as
\begin{equation}
P=\int_{-\infty}^\infty  \int_{-\infty}^\infty |u(x, y)|^2 dxdy.
\end{equation}
All the three power curves are non-monotonic. They have non-zero
minimum values inside the bandgap, below which solitary waves of the
respective family do not exist. This contrasts the 1D case where
solitary waves exist at all power levels \cite{Peli_1D}. Of these
three power curves, the one for the single-Bloch-mode family is the
lowest. Thus single-Bloch-mode solitons take the least power amount
to excite. Solitary waves of all three families at $\mu=7.189$ and
6.189, which are 0.04 and 1.04 below the edge point C (where
$\mu_0=7.229$), are plotted in Fig. \ref{soliton_C}. Note that the
shorter separation $\mu-\mu_0=-0.04$ is chosen so that
$\mu-\mu_0=-\epsilon^2$, where $\epsilon=0.2$ is the value we have
used when plotting all analytical solutions $u_0(x, y)$ in Figs.
\ref{soliton_C_anal0} and \ref{soliton_C_anal}. The solitons at
$\mu=7.189$, shown in Fig. \ref{soliton_C}(a, c, f), are weakly
localized and have low amplitudes, and they are close to the band
edge C. These numerical solutions are almost identical to the
analytical ones shown in Figs. \ref{soliton_C_anal0}(b),
\ref{soliton_C_anal}(c) and \ref{soliton_C_anal} (e) (the phase
fields of solitons in Fig. \ref{soliton_C}(c, f) are also almost
identical to the analytical ones in Fig. \ref{soliton_C_anal}(d, f)
and thus not shown). Thus the numerical results corroborate the
analytical theory of the previous section. More significant
solutions in Fig. \ref{soliton_C} are the three solitons at the
other propagation constant $\mu=6.189$, which is deep inside the
first bandgap. These solutions, displayed in Fig.
\ref{soliton_C}(b), (d, e) and (g, h), are strongly localized.
Indeed, the soliton of the single-Bloch-mode family in Fig.
\ref{soliton_C}(b) almost becomes a single dipole aligned along the
$y$ axis at the lattice site of origin $(x, y)=(0,0)$; the soliton
of the dipole-array family in Fig. \ref{soliton_C}(d, e) almost
becomes a single dipole aligned along the $y=-x$ direction at the
lattice site of origin; and the soliton of the vortex-array family
in Fig. \ref{soliton_C}(g, h) almost becomes a single vortex at the
lattice site of origin. The phase structures of these strongly
localized solitons, however, still resemble those of weakly
localized ones (see Figs. \ref{soliton_C_anal}(d, f) and
\ref{soliton_C}(e, h)). This strong localization of solitons in
these solution families makes them amiable for on-axis excitation in
photorefractive crystals. Indeed, the soliton of vortex-array family
in Fig. \ref{soliton_C}(g, h) looks almost identical to the gap
vortex soliton observed in \cite{Segev_highervortex} with on-axis
excitation. It should be noted that reduced-symmetry solitons
observed in \cite{Kivshar_C_onemode} also bifurcate from single
Bloch modes at point C. However, such solitons observed in
\cite{Kivshar_C_onemode} at high powers look a little different from
that in Fig. \ref{soliton_C}(b): their solitons are confined to
three lattice sites, with the middle-site peak intensity higher than
those at the two neighboring sites; while the soliton of Fig.
\ref{soliton_C}(b) is confined to two lattice sites with equal peak
intensities. The reason for this difference is the following. As we
have explained in Sec. \ref{envelope_derivation}(B), envelopes of
Bloch modes can be centered at four locations (see Eq.
(\ref{locations})). For the single-Bloch-mode soliton family shown
in Figs. \ref{powerC} and \ref{soliton_C}(a, b), the envelope was
centered at the origin $(x_0, y_0)=(0, 0)$, which leads to a dipole
structure under strong localization. If the envelope is centered at
$(x_0, y_0)=(0, L/2)=(0, \pi/2)$ instead, solitons under strong
localizations would have a central peak shouldered by two equal but
lower intensity peaks just like those observed in
\cite{Kivshar_C_onemode}. Thus, solitons observed in
\cite{Kivshar_C_onemode} belong to the single-Bloch-mode soliton
family of point C with the envelope centered at $(0, \pi/2)$ rather
than $(0, 0)$. From the experimental point of view, dipole-array
solitons as shown in Fig. \ref{soliton_C}(c,d,e), as well as the
strongly localized single-Bloch-mode soliton of dipole-type in Fig.
\ref{soliton_C}(b), have never been observed before, and they still
await experimental demonstration.

\subsection{Solution families bifurcated from point D}

\begin{figure}
\center
\includegraphics[width=0.45\textwidth]{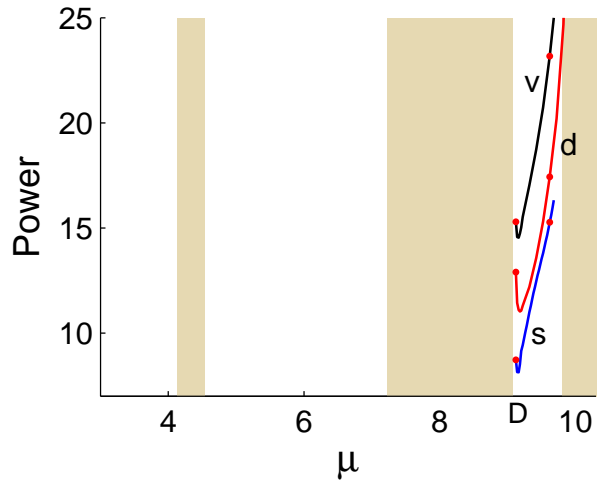}
\caption{(color online) Power diagrams of the three families of
solitary waves bifurcated from the edge point D (under defocusing
nonlinearity). Bottom curve: single-Bloch-mode branch (marked by
letter 's'); middle curve: dipole-cell branch (marked by letter
'd'); top curve: vortex-cell branch (marked by letter 'v').  The
marked thick dot points are 0.04 and 0.54 above the band edge D.
\label{powerD}}
\end{figure}

\begin{figure}
\center
\includegraphics[width=0.4\textwidth]{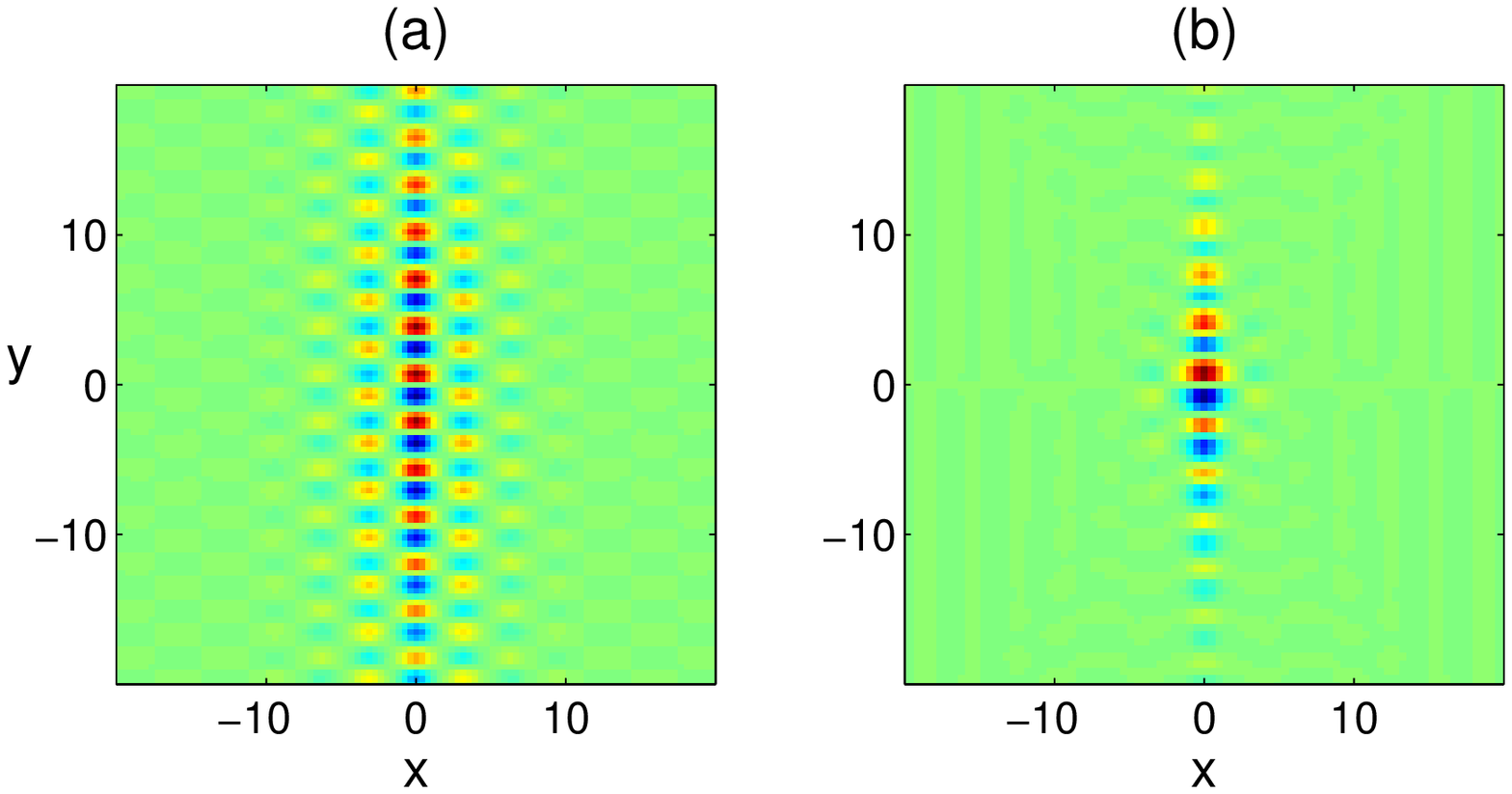}

\includegraphics[width=0.6\textwidth]{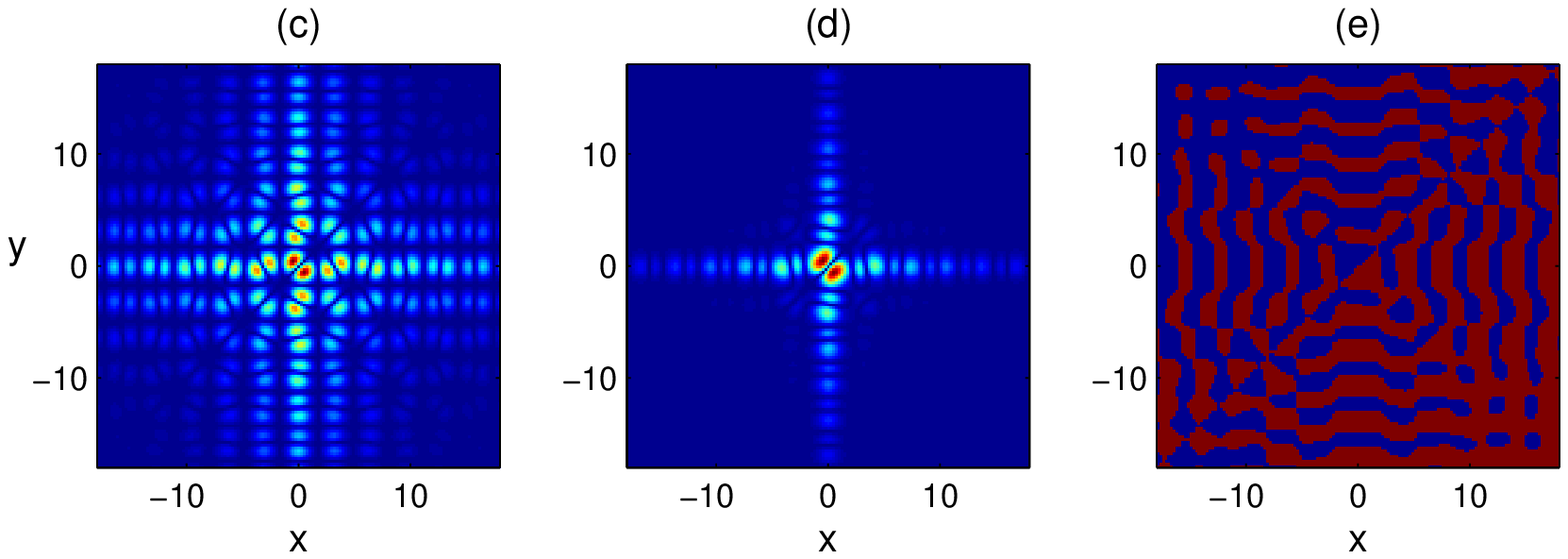}

\includegraphics[width=0.6\textwidth]{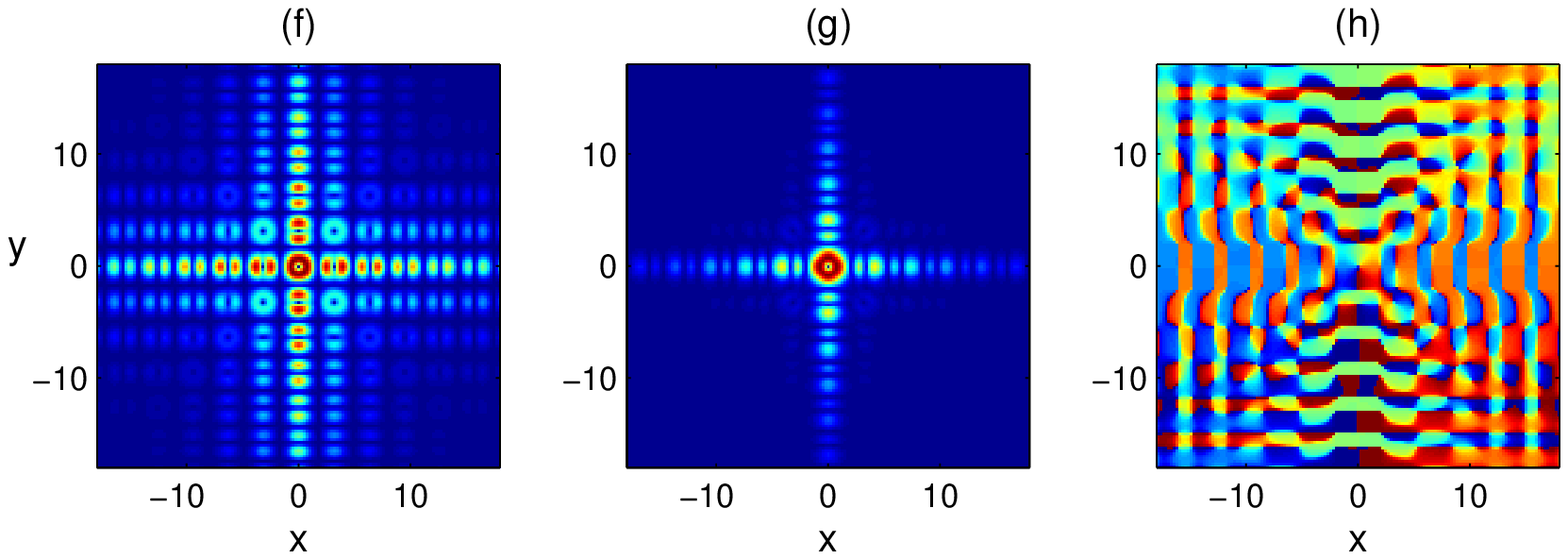}
\caption{(color online) Weakly and strongly localized solutions on
the three soliton branches bifurcated from the edge point D in Fig.
\ref{powerD}. Propagation constants of these solutions are
$\mu=9.121$ and 9.621, which are marked by thick dots in Fig.
\ref{powerD}. Top row: single Bloch-mode solitons; $u(x, y)$ is
shown; (a) $\mu=9.121$; (b) $\mu=9.621$. Middle and bottom rows:
dipole-cell and vortex-cell solitons respectively; (c, f) amplitudes
($|u(x, y)|$) at $\mu=9.121$; (d, e) amplitude and phase of the
dipole-cell soliton at $\mu=9.621$; (g, h) amplitude and phase of
the vortex-cell soliton at $\mu=9.621$.  \label{soliton_D}}
\end{figure}

At point D, three types of solitary waves were predicted in Sec.
\ref{envelope_section} under {\it defocusing} nonlinearity:
single-Bloch-mode solitons, dipole-cell solitons, and vortex-cell
solitons, see Figs. \ref{soliton_D_anal0}(b),
\ref{soliton_D_anal}(c, d) and \ref{soliton_D_anal} (e, f)
respectively. They all exist in the second bandgap, bifurcated from
the edge point D of the second Bloch band. Numerically, we have
obtained the entire families of these three types of solutions.
Their power curves are displayed in Fig. \ref{powerD}. These power
curves are also non-monotonic and have non-zero minimum values
inside the bandgap. Solitary waves of these three families at
$\mu=9.121$ and 9.621, which are 0.04 and 0.54 above the edge point
D (where $\mu_0=9.081$), are plotted in Fig. \ref{soliton_D}. The
solitons at $\mu=9.121$, shown in Fig. \ref{soliton_D}(a, c, f), are
weakly localized and have low amplitudes. They are located close to
the band edge D. These numerical solutions are almost
indistinguishable from the analytical ones shown in Figs.
\ref{soliton_D_anal0}(b), \ref{soliton_D_anal}(c) and
\ref{soliton_D_anal} (e) where $\epsilon=0.2$ (the phase fields of
solitons in Fig. \ref{soliton_D}(c, f) are also almost
indistinguishable from the analytical ones in Fig.
\ref{soliton_D_anal}(d, f) and thus not shown), thus numerical and
analytical solutions near the band edge D are in agreement.
Solutions at $\mu=9.621$ away from the edge point D are more
localized. These solutions are displayed in Fig. \ref{soliton_D}(b),
(d, e) and (g, h) respectively. The single-Bloch-mode soliton in
Fig. \ref{soliton_D}(b) is confined to only one lattice site along
the $x$ direction, but occupies quite a few lattice sites along the
$y$ direction. Thus this soliton will be highly mobile along the $y$
direction and strongly trapped along the $x$ direction. The
dipole-cell soliton in Fig. \ref{soliton_D}(d, e) is largely
confined to the single lattice site at the origin as a dipole
aligned along the $y=-x$ direction, with long tails stretching along
the horizontal and vertical (i.e. lattice) directions in a cross
pattern. The vortex-cell soliton in Fig. \ref{soliton_D}(g, h) is
also largely confined to the single lattice site at the origin in
the form of a vortex-ring with charge $-1$, with long tails along
the horizontal and vertical axes as well. None of these solitary
wave structures in Fig. \ref{soliton_D} was theoretically predicted
or experimentally observed before. These structures exist under
defocusing nonlinearity. Such nonlinearity can be obtained in
photorefractive crystals with a negative bias charge or in
Bose-Einstein condensates for certain types of atoms such as
$^{87}$Rb and $^{23}$Na \cite{Pitaevskii}.

\subsection{Solution families bifurcated from point E}

\begin{figure}
\center
\includegraphics[width=0.45\textwidth]{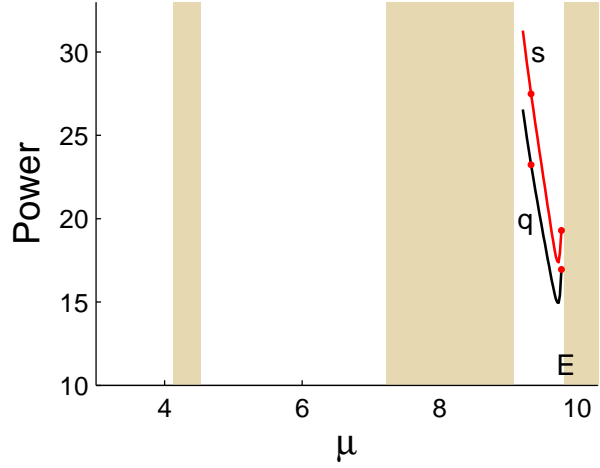}
\caption{(color online) Power diagrams of the two families of
solitary waves bifurcated from the edge point E (under focusing
nonlinearity). Upper curve: the spike-array branch (marked by letter
's'); lower curve: the quadrupole-array branch (marked by letter
'q'). The marked thick dot points are 0.04 and 0.48 below the band
edge E. \label{powerE}}
\end{figure}

\begin{figure}
\center
\includegraphics[width=0.45\textwidth]{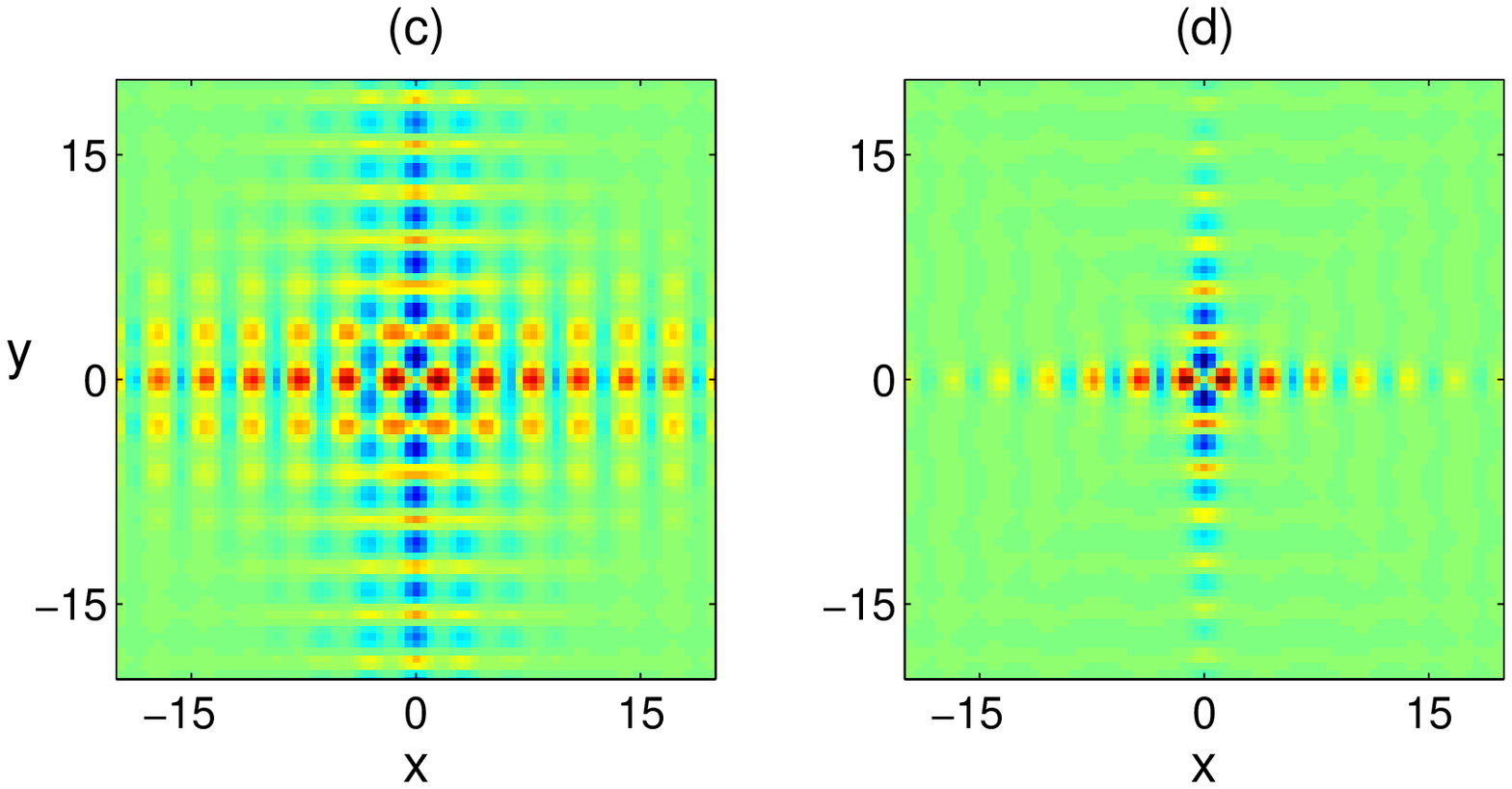}

\includegraphics[width=0.45\textwidth]{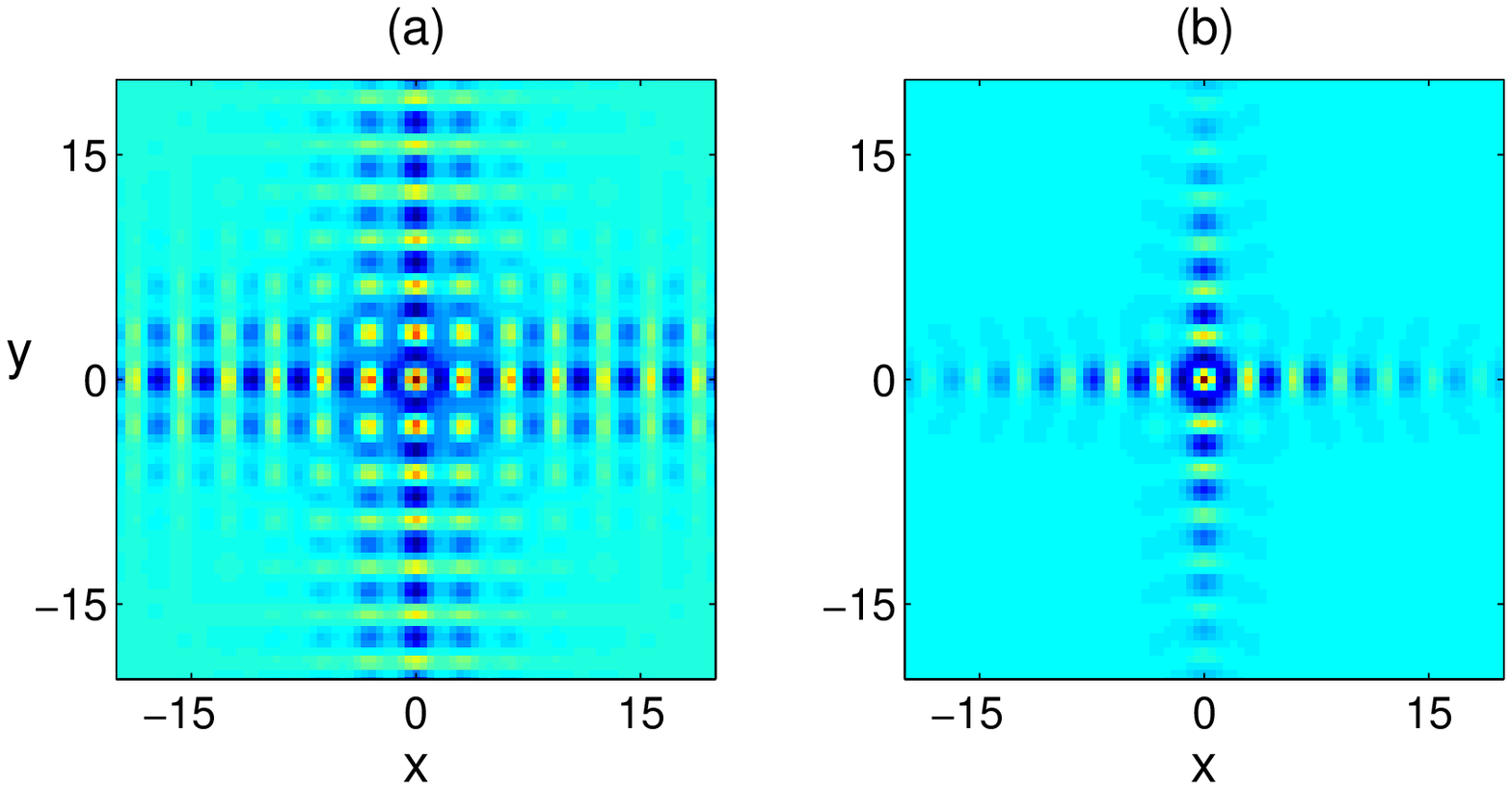}
\caption{(color online) Weakly and strongly localized solutions on
the two soliton branches bifurcated from the edge point E in Fig.
\ref{powerE}. Upper row: spike-array solitons; lower row:
quadrupole-array solitons. Propagation constants of these solutions
are $\mu=9.77$ (left column) and 9.33 (right column), which are
marked by thick dots in Fig. \ref{powerE}. \label{soliton_E}}
\end{figure}

At point E, two types of solitary waves were predicted in Sec.
\ref{envelope_section} under {\it focusing} nonlinearity:
spike-array solitons and quadrupole-array solitons, see Fig.
\ref{soliton_E_anal}(c, d). They exist in the second bandgap,
bifurcated from the edge point E of the third Bloch band.
Numerically, we have obtained families of these two types of
solutions. Their power curves are displayed in Fig. \ref{powerE}.
These power curves are also non-monotonic and have non-zero minimum
values inside the bandgap. Solitary waves of these families at
$\mu=9.77$ and 9.33, which are 0.04 and 0.48 above the edge point E
(where $\mu_0=9.81$), are plotted in Fig. \ref{soliton_E}. The
solitons at $\mu=9.77$, shown in Fig. \ref{soliton_E}(a, c), are
weakly localized and have low amplitudes. They are almost the same
as the analytical solutions in Fig. \ref{soliton_E_anal}(c, d) where
$\epsilon=0.2$, which is expected. The solutions at $\mu=9.33$ away
from the edge point E are more localized. These solutions are
displayed in Fig. \ref{soliton_E}(b, d). The localized solution of
the spike-array soliton family in Fig. \ref{soliton_E}(b) is now
confined to only a few lattice sites on the $x$ and $y$ axes, with a
dominant circular spike in its center. The localized solution of the
quadrupole-array soliton family in Fig. \ref{soliton_E}(d) is
confined also to only a few lattice sites on the $x$ and $y$ axes,
with a quadrupole in its center. These types of solitary waves are
novel and have not been reported before.

\section{Time-dependent envelope equations}

In the previous sections, our focus was to obtain solitary waves in
Eq. (\ref{NLS}), thus our envelope equations (\ref{A})-(\ref{B}) and
(\ref{A_single}) were time-independent. If we look beyond solitary
waves, we can further study how slowly modulated low-amplitude
Bloch-wave packets evolve with time under nonlinear effects. In such
a study, envelope equations of Bloch modes will be time-dependent.
Such time-dependent envelope equations are essential not only for
the tracking of Bloch-wave packets' movement, but also for the
stability analysis of stationary envelope solutions obtained in Sec.
\ref{envelope_section}. This is analogous to the role the nonlinear
Schr\"odinger equation played in the study of water wave packets and
light pulses in nonlinear optics \cite{Benney_Roskes,
Hasegawa_Kodama}. Derivation of time-dependent envelope equations is
a simple extension of the analysis in Sec.
\ref{envelope_derivation}, and will be performed briefly below. As
before, the derivation will be carried out for the slightly more
complicated case where the solution is near band edges with two
different Bloch modes. Results for the simpler case of the solution
being near band edges with a single Bloch mode will be given without
elaboration.

Consider the evolution of low-amplitude Bloch-wave packets in Eq.
(\ref{NLS}) near a band edge $\mu=\mu_0$ with two different Bloch
modes $p_1(x)p_2(y)$ and $p_2(x)p_1(y)$. The solution can be
expanded into the following perturbation series:
\begin{equation} \label{Uexpand}
U(x, y, t)=e^{-i\mu_0 t}\left(\epsilon U_0+\epsilon^2 U_1+
\epsilon^3 U_2 +\dots\right),
\end{equation}
where
\begin{equation}
U_0=A_1(X,Y,T)p_1(x)p_2(y)+A_2(X,Y,T)p_2(x)p_1(y),
\end{equation}
$\e \ll 1$, $X=\e x, Y=\e y$ are slow spatial variables, and
$T=\epsilon^2 t$ is the slow time variable. When this expansion is
substituted into Eq. (\ref{NLS}), we find at $O(\e^2)$ that the
equation for $U_1$ is simply Eq. (\ref{u1}) with $u_k \: (k=0, 1)$
replaced by $U_k$, and with an additional term $i\partial
U_1/\partial t$ added onto its left hand side. The solution for
$U_1$ is still given by Eq. (\ref{u1_solution}). At $O(\e^3)$, we
find that the equation for $U_2$ is Eq. (\ref{u2}) with $u_k \:
(k=0, 1, 2)$ replaced by $U_k$, with an additional term $i\partial
U_2/\partial t$ added onto its left hand side, and with $\eta u_0$
replaced by $i\partial U_0/\partial T$. Requiring the $U_2$ solution
not to grow linearly with time $t$ (suppression of secular terms),
the following time-dependent equations will be derived for the
envelope functions $A_1(X, Y, T)$ and $A_2(X, Y, T)$:
\begin{widetext}
\begin{eqnarray}
i\frac{\partial A_1}{\partial T} + D_1 \frac{\p^2A_1}{\p X^2}+D_2
\frac{\p^2A_1}{\p Y^2}+ \sigma \left[\alpha |A_1|^2A_1+\beta
\left(\bar{A}_1A_2^2+2 A_1|A_2|^2\right)+\gamma\left(|A_2|^2A_2+\bar{A}_2A_1^2+2A_2|A_1|^2\right)\right]=0,    \label{A1T} \\
i\frac{\partial A_2}{\partial T} + D_2 \frac{\p^2A_2}{\p
X^2}+D_1\frac{\p^2A_2}{\p Y^2} + \sigma
\left[\alpha|A_2|^2A_2+\beta\left(\bar{A}_2A_1^2+2 A_2|A_1|^2\right)
+\gamma\left(|A_1|^2A_1+\bar{A}_1A_2^2+2A_1|A_2|^2\right)\right]=0.
\label{A2T}
\end{eqnarray}
\end{widetext}
Here all coefficients $D_1, D_2, \alpha, \beta$ and $\gamma$ are the
same as those given in Eqs. (\ref{Dn}) and
(\ref{alpha})-(\ref{gamma}).

If the solution of Eq. (\ref{NLS}) is a low-amplitude Bloch-wave
packet near a band edge $\mu=\mu_0$ where a single Bloch mode
$p_1(x)p_1(y)$ exists, the perturbation expansion for the solution
$U$ is still Eq. (\ref{Uexpand}), except that
\begin{equation}
U_0=A_1(X,Y,T) p_1(x)p_1(y)
\end{equation}
now. In this case, the equation for the envelope function
$A_1(X,Y,T)$ reduces to
\begin{equation}  \label{A1T_single}
i\frac{\partial A_1}{\partial T} + D_1 \left(\frac{\p^2A_1}{\p X^2}+
\frac{\p^2A_1}{\p Y^2}\right)+ \sigma \alpha_0 |A_1|^2A_1=0,
\end{equation}
where the coefficients $D_1$ and $\alpha_0$ are given in Eqs.
(\ref{Dn}) and (\ref{alpha0}).

Using the above time-dependent envelope equations, one can study the
time evolution of Bloch-wave packets. In addition, one can analyze
linear and nonlinear stabilities of stationary Bloch-wave packet
solutions (i.e. solitary waves in Eqs. (\ref{A1T})-(\ref{A2T}) and
(\ref{A1T_single})). Such studies lie outside the scope of the
present article.


\section{Summary and Discussion}
In this paper, we systematically studied various families of
solitary waves which are bifurcated from Bloch-band edges in the
two-dimensional NLS equation with a periodic potential. Near the
band edges, we analytically derived envelope equations for
low-amplitude Bloch-wave packets. Based on these envelope equations,
many novel types of solitary waves inside bandgaps were predicted,
including dipole-array solitons, dipole-cell solitons, vortex-cell
solitons, etc. Away from the band edges, solitary waves were traced
numerically, and strongly localized solitary waves of respective
families were obtained. These solitary waves have distinctive
intensity and phase patterns such as cross-shape intensity
distributions and vortex-cell phase profiles. Certain solitary waves
previously reported in 2D photonic lattices such as reduced-symmetry
solitons \cite{Kivshar_C_onemode} and gap vortex solitons
\cite{Segev_highervortex} were found to be special cases of our
general classes of solutions.

In this paper, the relatively simple and obvious types of solutions
in envelope equations were determined, and their induced solitary
wave families computed. Weather other solutions of envelope
equations would generate additional families of solitary waves is
still an open question. For instance, the single-envelope equation
(\ref{A_single}) also admits vortex-ring solutions of the type
$f(R)e^{in\Theta}$. Whether such solutions would lead to new
families of vortex-type lattice solitons remains to be seen. For the
coupled envelope equations (\ref{A})-(\ref{B}), it is not even clear
what kind of additional solutions they admit beyond the ones we
presented in this article. Thus a classification of solutions in the
coupled envelope equations (\ref{A})-(\ref{B}) is highly desirable.
Linear and nonlinear stabilities of solitary waves reported in this
paper is another open question which merits careful investigation.
From the experimental point of view, the challenge is to
experimentally demonstrate the new types of solitary waves obtained
in this article, such as dipole-array solitons and vortex-cell
solitons. These open questions are beyond the scope of the present
article, and will be left for future studies.

\section*{Acknowledgments}
The authors appreciate helpful discussions with Dr. Zhigang Chen.
This work was partially supported by the Air Force Office of
Scientific Research under grant USAF 9550-05-1-0379.




\end{document}